\documentclass[aps,prc,10pt,showpacs,showkeys,twocolumn,superscriptaddress,groupedaddress]{revtex4-2}
\usepackage[latin1]{inputenc}
\usepackage[english]{babel}
\usepackage{graphicx,psfrag}
\usepackage{amsmath}
\usepackage{amssymb}
\usepackage{amscd}
\usepackage{eucal}
\usepackage{color}
\usepackage{bm}
\usepackage{braket}
\usepackage{dsfont}
\begin{document}
\title{{\it Ab initio} calculations of electric dipole moments of light nuclei}

\author{Paul Froese}
\email{pafroese@student.ubc.ca}
\affiliation{TRIUMF, 4004 Wesbrook Mall, Vancouver, British Columbia V6T 2A3, Canada}
\affiliation{Department of Physics and Astronomy, University of British Columbia, Vancouver, British Columbia V6T 1Z1, Canada}

\author{Petr Navr\'atil}
\email{navratil@triumf.ca}
\affiliation{TRIUMF, 4004 Wesbrook Mall, Vancouver, British Columbia V6T 2A3, Canada}

\date{\today}

\begin{abstract}
 In any finite system, the presence of a non-zero permanent electric dipole moment (EDM) would indicate CP violation beyond the small violation predicted in the Standard Model. Here, we use the {\it ab initio} no-core shell model (NCSM) framework to theoretically investigate the magnitude of the nuclear EDM. We calculate EDMs of several light nuclei using chiral two- and three-body interactions and a PT-violating Hamiltonian based on a one-meson-exchange model. We present a benchmark calculation for $^3$He, as well as results for the more complex nuclei $^{6,7}$Li, $^9$Be, $^{10,11}$B, $^{13}$C, $^{14,15}$N, and $^{19}$F.  Our results suggest that different nuclei can be used to probe different terms of the PT violating interaction. These calculations 
allow us to suggest which nuclei may be good candidates in the search for a measurable permanent electric dipole moment.
\end{abstract} 
\maketitle

\section{INTRODUCTION}

A permanent electric dipole moment (EDM) of a physical system would indicate direct violation of time-reversal (T) and parity (P) and thus charge conjugation and parity (CP) violation through the CPT invariance. CP violation is a required condition for baryogenesis in the early universe~\cite{Sakharov:1967dj}. In the Standard Model (SM) with three generations of quarks, CP is broken by the phase of the Cabibbo-Kobayashi-Maskawa (CKM) mixing matrix~\cite{CKM1973} and by the QCD $\bar{\theta}$ term~\cite{PhysRevLett.37.8}. While observed CP violation in the kaon and B meson systems can be explained by the CKM mechanism, CP violation in the SM fails to generate the observed matter-antimatter asymmetry of the Universe by several orders of magnitude~\cite{Gavela_1994a,Gavela_1994b}. 

The CKM mechanism predicts values for the EDMs of leptons, nucleons, atomic and molecular systems that are too small to be detected in the foreseeable future, and hence a measured nonzero EDM in any of these systems is an unambiguous signal for a new source of CP violation and for physics beyond the SM~\cite{RevModPhys.91.015001}. The present experimental upper bounds on the EDMs of neutron and proton are $|d_n|{<}1.8 \times 10^{-13} e$~fm~\cite{nEDM2020} and $|d_p|{<}2 \times 10^{-12} e$~fm, where the proton EDM has been inferred from a measurement of the diamagnetic $^{199}$Hg atom~\cite{Graner2016}. For the electron, the most recent upper bound is $|d_e|{<}8.7 \times 10^{-16} e$~fm~\cite{ACME2014}, derived from the EDM of the ThO molecule.

In this letter, we focus on nuclear EDMs. There are proposals to measure the EDMs of charged particles, including protons and light nuclei, in dedicated storage ring experiments~\cite{Orlov2006,Pretz_2013,Hempelmann2017,abusaif2019storage}. These experiments might reach a sensitivity of $10^{-16} e$~fm, comparable with the next generation of neutron EDM experiments. Unlike searches for CP-violating moments of the nucleus through measurements of atomic EDMs, a measurement for a stripped nucleus would not suffer from a suppression of the signal through atomic Schiff screening~\cite{Schiff1963}. In comparison to a proton or a neutron EDM, EDMs of atomic nuclei can be enhanced by many-body effects~\cite{Flambaum1985}.

EDMs of few nucleon systems, the deuteron, $^3$H, $^3$He, have been investigated by various {\it ab initio} approaches~\cite{Liu2004,STETCU2008168,deVries2011d,deVries2011,Lazauskas2013,Bsaisou2015,Wirzba2017,Gnech2020} using phenomenological meson-exchange and/or chiral Effective Field Theory (EFT) interactions as well as within pionless EFT framework~\cite{yang2020electric}. Recently, EDMs of selected $p$-shell nuclei were calculated within the cluster model~\cite{Yamanaka2015,Yamanaka2017a,Yamanaka2017b,Yamanaka2018,Yamanaka2019a,Yamanaka2019b}. In particular, EDMs were reported for $^6$Li~\cite{Yamanaka2015}, $^9$Be~\cite{Yamanaka2019a}, $^{7}$Li and $^{11}$B~\cite{Yamanaka2019b}, and $^{13}$C~\cite{Yamanaka2017a} using phenomenological cluster-cluster PT-conserving (PTC) interaction and one-meson-exchange based PT-violating (PTV) nucleon-nucleon (NN) interaction.

In this work, we perform {\it ab initio} calculations of EDMs for light nuclei within the no-core shell model (NCSM) ~\cite{PhysRevLett.84.5728,PhysRevC.62.054311,Barrett2013} framework using chiral NN and three-nucleon (3N) PTC interactions and one-meson-exchange PTV NN interactions as the only input. The NCSM is applicable in a universal way to few-nucleon systems, $p$-shell, and light $sd$-shell nuclei. We present benchmark calculations for $^3$He as well as results for the more complex stable nuclei $^{6,7}$Li, $^9$Be, $^{10,11}$B, $^{13}$C, $^{14,15}$N, and $^{19}$F. We note that NCSM was applied to obtain the first {\it ab initio} EDM results for $^3$He and $^3$H in Refs.~\cite{STETCU2008168} and ~\cite{deVries2011}, respectively.

\section{NO-CORE SHELL MODEL}

In the NCSM, nuclei are described as systems of $A$ non-relativistic point-like nucleons interacting through realistic inter-nucleon interactions. All nucleons are active degrees of freedom. The many-body wave function is cast into an expansion over a complete set of antisymmetric $A$-nucleon harmonic oscillator (HO) basis states containing up to  $N_{\rm max}$ HO excitations above the lowest Pauli-principle-allowed configuration. 
The basis is further characterized by the frequency $\Omega$ of the HO well. Square-integrable energy eigenstates  are obtained by solving the Schr\"{o}dinger equation
\begin{equation}\label{Sch_eq}
H \ket{A\, \lambda\; I^\pi } = E_\lambda ^{I^\pi } \ket{A\, \lambda\; I^\pi } \; ,
\end{equation}
with the intrinsic PTC Hamiltonian
\begin{equation}\label{eq:ham_Ham}
H=\frac{1}{A}\sum_{i<j=1}^A\frac{(\vec{p}_i-\vec{p}_j)^2}{2m} + \sum_{i<j=1}^A V^{\rm NN}_{ij} + \sum_{i<j<k=1}^A V^{\rm 3N}_{ijk} \, ,
\end{equation}
Here, $m$ is the nucleon mass, $\vec{p}$ nucleon momenta, $V^{\rm NN}$ and $V^{\rm 3N}$ PTC NN and 3N interaction, respectively. The $\lambda$ in~\eqref{Sch_eq} labels eigenstates with identical $I^\pi$. The eigenstates of $H$~(\ref{eq:ham_Ham}) can be also characterized by isospin quantum number $T$ that is typically conserved to a good approximation. We note, however, that our calculations fully include isospin breaking originating from the Coulomb interaction and strong force contributions present in the $V^{\rm NN}$.

The present calculations are performed using the Slater determinant (SD) HO basis in the so-called $M$-scheme where the basis is characterized by $A$, the projection $I_z$ of the total angular moment $I$, parity $\pi$ and $T_z{=}(Z{-}N)/2$ with $Z$ and $N$ the proton and neutron number, respectively. Only the eigenstates~\eqref{Sch_eq} obtained by diagonalization using the Lanczos algorithm have good $I$ and approximately good $T$. They factorize exactly as products of physical intrinsic eigenstates and a center-of-mass state in the $0\hbar\Omega$ excitation.

In the present work we adopt the NN+3N chiral interaction applied in Ref.~\cite{PhysRevC.101.014318}, denoted as NN+3N(lnl), consisting of an NN interaction up to the fourth order (N$^3$LO) in the chiral expansion~\cite{Entem2003} and a 3N interaction up to next-to-next-to-leading order (N$^2$LO) using a combination of local and non-local regulators. Even though all the underlying parameters (known as low-energy constants or LECs) are determined in $A{=}2,3,4$ nucleon systems, this interaction provides a very good description of properties of both light and medium mass nuclei~\cite{PhysRevC.101.014318}, including $^{100}$Sn~\cite{Gysbers2019}. The chiral orders of the adopted NN and 3N interactions are not consistent: the former is included up to order N$^3$LO while the latter is at N$^2$LO. While the N$^3$LO 3N contribution has been shown to be rather small~\cite{Machleidt2011}, the consistency of the regulator and/or in particular the use of a non-local versus local regulators plays a significant role in medium mass nuclei~\cite{Huther2020}.

A faster convergence of our calculations with respect to the many-body basis size is obtained by softening the chiral interaction through the similarity renormalization group (SRG) technique~\cite{Wegner1994,Bogner2007,PhysRevC.77.064003,Bogner201094,Jurgenson2009}. The SRG unitary transformation induces many-body forces, included here up to the three-body level. The four- and higher-body induced terms are small at the $\Lambda_{\mathrm{SRG}}{=}2.0$ fm$^{-1}$ resolution scale used in present calculations~\cite{PhysRevC.101.014318}.

\section{THE NUCLEAR ELECTRIC DIPOLE MOMENT}

A nuclear EDM consists of contributions from the intrinsic EDMs of the proton and neutron, $d_p$ and $d_n$ and from the polarization effect caused by the PTV nuclear interaction, as well as from the two-body PTV meson-exchange charge operator. The latter was found to be just a few percent  of the polarization contribution for the deuteron case~\cite{Liu2004} and will not be considered in this work.

Contributions due to intrinsic EDMs of the nucleons can be evaluated by calculating the matrix element
\begin{eqnarray}\label{eq:D1}
  D^{(1)} &=& \bra{A\, {\rm gs}\; I^\pi I_z{=}I}  \\ \nonumber
  &\times& \sum_{i=1}^A \frac{1}{2}[(d_p+d_n)+(d_p-d_n)\tau_{i,z}]\sigma_{i,z} \\ \nonumber
  &\times& \ket{A\, {\rm gs}\; I^\pi I_z{=}I} \; ,
\end{eqnarray}
where the ground state wave function is obtained by solving the Schr\"odinger equation (\ref{Sch_eq}) with the PTC Hamiltonian (\ref{eq:ham_Ham}). The $\tau$ and $\sigma$ are nucleon isospin and spin operators, respectively. 

The PTV NN interaction admixes unnatural parity states in the ground state
\begin{eqnarray}\label{gswf}
  |A \, {\rm gs}\; I \rangle &=& |A \, {\rm gs}\; I^\pi \rangle + \sum_\lambda  |A\, \lambda \; I^{-\pi}\rangle \\ \nonumber
                                 &\times& \frac{1}{E_{\rm gs}^{I^\pi}-E_\lambda^{I^{-\pi}}} \langle A\, \lambda \; I^{-\pi}| V_{\rm NN}^{\rm PTV}|A \, {\rm gs} \; I^\pi \rangle \; ,
\end{eqnarray}  
which then gives rise to the induced EDM moment. We use the one-meson-exchange model for the PTV NN interaction including the $\pi-$, $\rho-$, and $\omega-$meson exchanges in the form~\cite{Liu2004,Haxton83,Gudkov93}
\begin{eqnarray}\label{eq:PTVNN}
  V_{\rm NN}^{\rm PTV}&=&\frac{1}{2m}\{\mathbf{\sigma}_{-} \cdot \mathbf{\nabla} (-\bar{G}^0_\omega y_\omega(r))  \\
                      &+&\mathbf{\tau}_1\cdot\mathbf{\tau}_2 \, \mathbf{\sigma}_- \cdot \mathbf{\nabla} (\bar{G}^0_\pi y_\pi(r) -\bar{G}^0_\rho y_\rho(r)) \nonumber \\
                     &+&\frac{1}{2}\,\tau^z_+ \, \mathbf{\sigma}_- \cdot \mathbf{\nabla} (\bar{G}^1_\pi y_\pi(r) -\bar{G}^1_\rho y_\rho(r) -\bar{G}^1_\omega y_\omega(r)) \nonumber \\
                      &+&\frac{1}{2}\,\tau^z_- \, \mathbf{\sigma}_+ \cdot \mathbf{\nabla} (\bar{G}^1_\pi y_\pi(r) +\bar{G}^1_\rho y_\rho(r) -\bar{G}^1_\omega y_\omega(r)) \nonumber \\
   &+&(3 \tau^z_1 \tau^z_2-\mathbf{\tau}_1\cdot\mathbf{\tau}_2) \, \mathbf{\sigma}_- \cdot \mathbf{\nabla} (\bar{G}^2_\pi y_\pi(r) -\bar{G}^2_\rho y_\rho(r)) \} \;, \nonumber
\end{eqnarray}
where $\bar{G}^T_\chi{=}\bar{g}_\chi \, g_{\chi {\rm NN}}$ is a product of a PTV $\chi$-meson-nucleon coupling and its associate strong one, $y_\chi(r){=}e^{-m_\chi r}/(4\pi r)$ is the Yukawa function with a range determined by the mass of the exchanged $\chi$-meson, $\vec{r}{=}\vec{r}_1-\vec{r}_2$, $\vec{\sigma}_\pm{=}\vec{\sigma}_1\pm \vec{\sigma}_2$, and  $\vec{\tau}_\pm{=}\vec{\tau}_1\pm \vec{\tau}_2$.

In the NCSM, when the $\ket{A \, {\rm gs} \; I^\pi}$ is calculated in $N_{\rm max}$ space, the corresponding unnatural parity states appearing in Eq. (\ref{gswf}) are obtained in $N_{\rm max}{+}1$ space. It is not necessary to compute many excited unnatural parity states as Eq.~(\ref{gswf}) suggests. Rather, first, we solve the standard Schr\"{o}dinger equation (\ref{Sch_eq}) using the PTC Hamiltonian (\ref{eq:ham_Ham}) and obtain the $\ket{A \, {\rm gs} \; I^\pi}$ wave function, and second, we invert the generalized Schr\"{o}dinger equation with an inhomogeneous term,
\begin{equation}\label{inhomeq}
  (E_{\rm gs}^{I^\pi}-H)  |A\, {\rm gs}\; I \rangle =  V_{\rm NN}^{\rm PTV}|A\, {\rm gs} \; I^\pi \rangle \; ,
\end{equation}
to obtain the unnatural parity admixture in the ground state. The inversion is performed by the Lanczos continued fraction method~\cite{Haydock_1974,Marchisio2003,STETCU2008168}.

The polarization contribution to the nuclear EDM is then calculated as
\begin{eqnarray}\label{eq:D2}
  D^{(\rm pol)}&=&\bra{A \, {\rm gs} \; I^\pi I_z{=}I} \frac{e}{2}\sum_{i=1}^A(1+\tau_i^z)z_i \ket{A \, {\rm gs} \; I  I_z{=}I} \nonumber \\
                   &+& {\rm h.c.}
\end{eqnarray}
with the electric dipole moment operator projected in the $z$-direction. With this form of the transition operator the leading effects of two-body electromagnetic currents are included through the Siegert theorem.

\section{RESULTS AND DISCUSSION}

To compute matrix elements of the $ V_{\rm NN}^{\rm PTV}$ interaction (\ref{eq:PTVNN}) and solve the equation (\ref{inhomeq}), we adapted codes used for calculations of anapole moments of light nuclei reported in Ref.~\cite{Hao2020}. To benchmark our codes, we calculated the EDM of $^3$He using PTC chiral N$^3$LO NN interaction~\cite{Entem2003} without any renormalization as $^3$He EDM results for this interaction together with the PTV interaction (\ref{eq:PTVNN}) were published in Ref.~\cite{STETCU2008168}.
\begin{figure}
\centering
\includegraphics[width=0.5\textwidth]{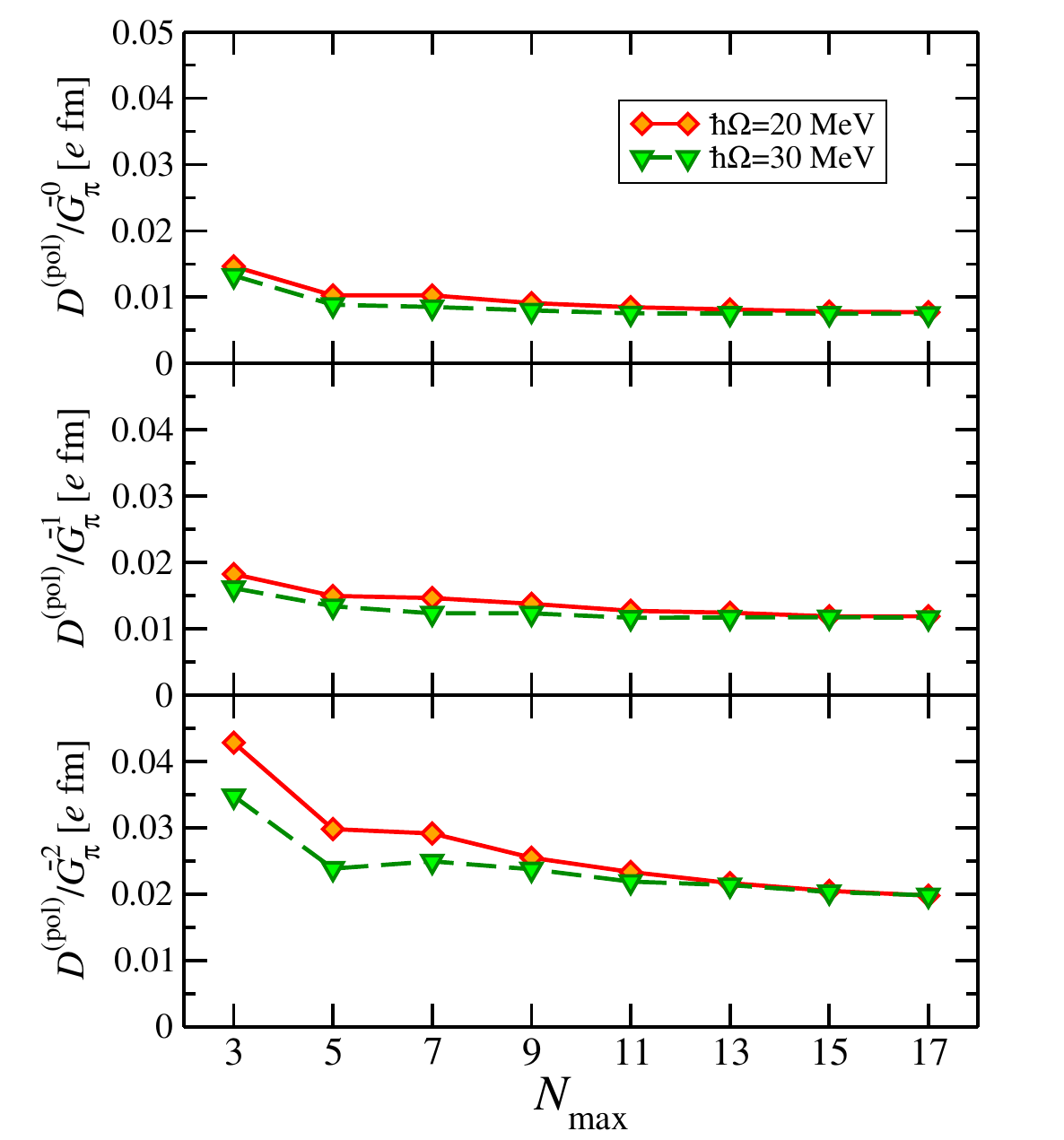}
\caption{\label{fig:3Heconv} The polarization contribution to $^3$He EDM (in $e$~fm) due to the $\pi$-exchange PTV NN interaction~(\ref{eq:PTVNN}). Dependence on the NCSM basis size characterized by $N_{\rm max}$ for two HO frequencies is shown. Chiral N$^3$LO PTC NN interaction from Ref.~\cite{Entem2003} was used.}
\end{figure}
The NCSM basis convergence for the polarization contribution to $^3$He EDM is shown in Fig.~\ref{fig:3Heconv} and our $D^{(1)}$ and $D^{({\rm pol})}$ results are summarized in Table~\ref{tab1}. The $D^{({\rm pol})}$ $N_{\rm max}$ convergence is quite satisfactory while that of $D^{(1)}$ is still faster. In Fig.~\ref{fig:3Heconv}, the odd $N_{\rm max}$ values correspond to the unnatural states in Eq.~(\ref{gswf}), i.e., the largest space for the ground-state was $N_{\rm max}{=}16$. While our $D^{(1)}$ results agree with those reported in Ref.~\cite{STETCU2008168} (Table 1, the EFT NN column in that paper), the present $D^{({\rm pol})}$ results are smaller by a factor of $1/2$ compared to Ref.~\cite{STETCU2008168} (Table 2, the EFT NN columns in that paper). It should be noted that the same $1/2$ discrepancy was reported in Ref.~\cite{Lazauskas2013} for the isoscalar and isovector terms, while a discrepancy of $1/5$ was found for the isotensor terms. Similarly, a factor of $1/2$ difference was found in Ref.~\cite{Yamanaka2015} although for all the terms. Our results are then consistent with those of Ref.~\cite{Yamanaka2015}. The NCSM was applied in Ref.~\cite{STETCU2008168} (and also in Ref.~\cite{deVries2011}). However, the Jacobi-coordinate HO basis was employed as opposed to the SD HO basis used here, i.e., different codes were utilized. We plan to reexamine the codes used in Ref.~\cite{STETCU2008168} to investigate the issue further.
 \begin{table*}[tb]
   \begin{center}
     \begin{ruledtabular}
       \begin{tabular}{l|cccccccccccc}
 & $d_p$ &  $d_n$ & $\bar{G}^0_\pi$ & $\bar{G}^1_\pi$ & $\bar{G}^2_\pi$ & $\bar{G}^0_\rho$ & $\bar{G}^1_\rho$ & $\bar{G}^2_\rho$ & $\bar{G}^0_\omega$ & $\bar{G}^1_\omega$ &$\mu$& $\mu^{\rm exp.}$ \\
         $^3$He  & -0.031   & 0.905  &  0.0073     & 0.011  &   0.019    & -0.00062    & 0.000063   & -0.0014   & 0.00042      & -0.00086 & -1.79 & -2.127 \\
         $^6$Li   & 0.892     & 0.890  &  0.00006    & 0.0171 &  0.0002   & -0.000003 & 0.00158     & -0.00002 & -0.000002  & -0.0016 & +0.84 & +0.822 \\
         $^7$Li   & 0.930     & 0.018  &  -0.0096  & 0.0106  & -0.0233  & 0.00131    & 0.00085   &  0.0029  & -0.00072  & -0.0013  & +2.99 & +3.256 \\
         $^9$Be  & 0.018     & 0.720  &     0.0007 &  0.0116 &  0.0053   & 0.00019    & 0.00005    & -0.0002 & 0.00046    &  -0.0004 & -1.05 & -1.177 \\
         $^{10}$B & 0.852    & 0.848  &  -0.0001  &  0.0281 &  -0.0002 & 0.00001    & 0.00075    &  0.00002  & -0.00002  &  -0.0017 & +1.83 & +1.801 \\
         $^{11}$B & 0.444    & 0.050  &   -0.0070 &  0.0127 & -0.0219  & 0.00039    & 0.00019    &   0.0019 & -0.00016   & -0.0010  & +2.09  & +2.689 \\
         $^{13}$C & -0.098  & -0.282 &  -0.0058 & -0.0084 & -0.0316 & 0.00016    & -0.00052  &   0.0037 &   0.00004   &  0.0010  & +0.44 & +0.702\\
         $^{14}$N & -0.366  & -0.363&    0.0003   & -0.0172 & 0.0006   & -0.00003  & -0.00081  & -0.0001 &  0.00002    &  0.0014 & +0.37 & +0.404\\
         $^{15}$N & -0.296  &   0.008 &    0.0102 & -0.0095 & 0.0228  & -0.00052   & -0.00044  &  -0.0015 &  0.00039   &  0.0008 & -0.25  & -0.283 \\
         $^{19}$F &  0.818   & -0.052  &  -0.0175 &   0.0089 & -0.0226 &  0.00236   & 0.00125    &   0.0027  & -0.00096  & -0.0014 & +2.85& +2.629 \\
       \end{tabular}
     \end{ruledtabular}
     \caption{The nucleonic and polarization contributions to EDMs of $^3$He, stable $p$-shell nuclei, and $^{19}$F (in $e$~fm) decomposed as coefficients of $d_p$, $d_n$, and $\bar{G}^T_\chi$, where $\chi$ stands for $\pi$, $\rho$, or $\omega$ exchanges. In the last two columns, calculated and experimental (from Ref.~\cite{STONE200575}) nuclear magnetic dipole moments (in $\mu_{\rm N}$) are compared. SRG-evolved chiral NN+3N(lnl) PTC interaction from Ref.~\cite{PhysRevC.101.014318} was used except for $^3$He where the chiral N$^3$LO PTC NN~\cite{Entem2003} was utilized.}
     \label{tab1}
   \end{center} 
\end{table*}
\begin{figure}
\centering
\includegraphics[width=0.49\textwidth]{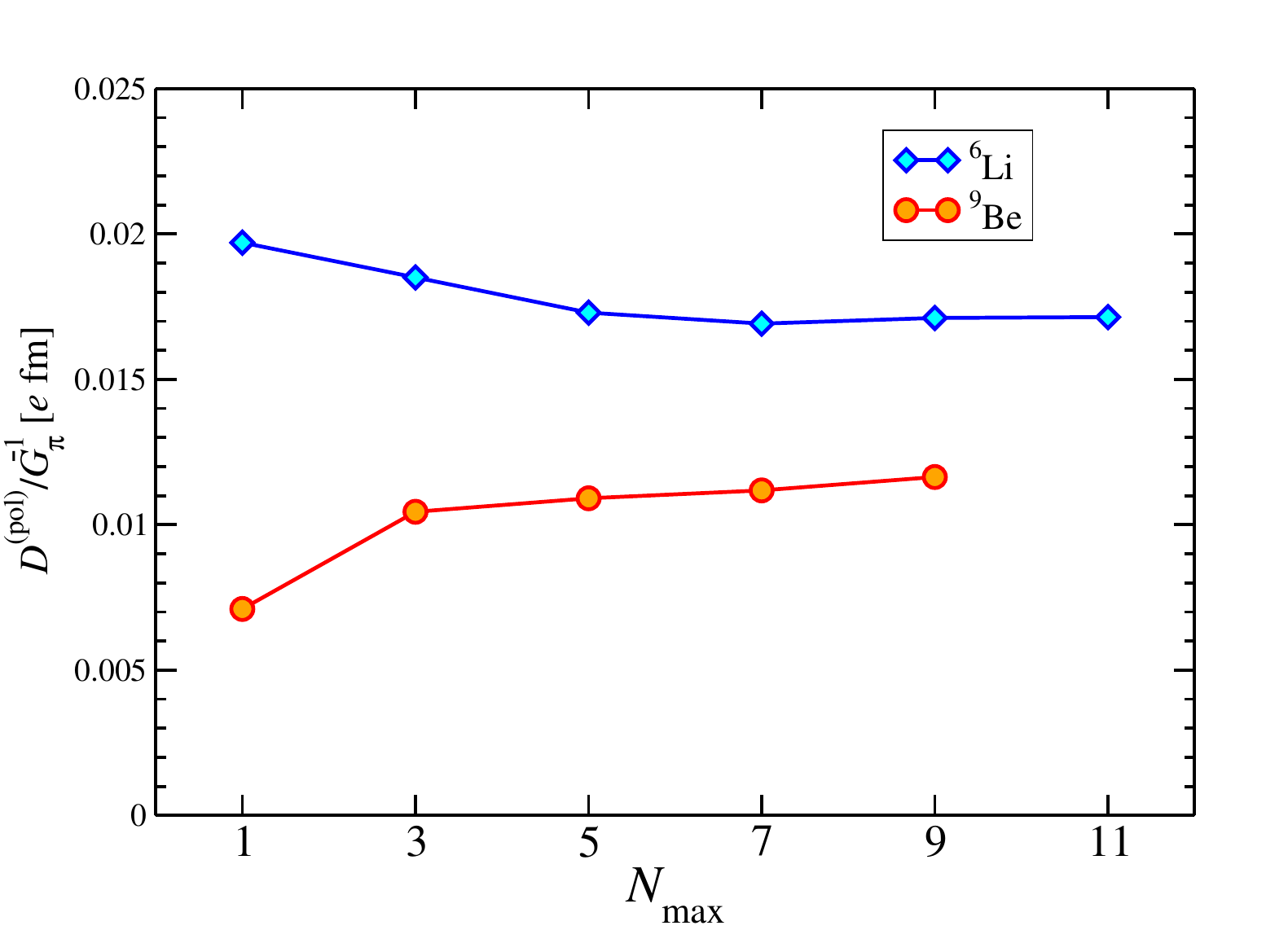}
\caption{\label{fig:LiBeconv} The polarization contribution to $^{6}$Li and $^9$Be EDM (in $e$ fm) due to the isovector $\pi$-exchange PTV NN interaction~(\ref{eq:PTVNN}). Dependence on the NCSM basis size characterized by $N_{\rm max}$ is shown. SRG-evolved chiral NN+3N(lnl) PTC interaction from Ref.~\cite{PhysRevC.101.014318} was used. The HO frequency $\hbar\Omega{=}20$ MeV was used.}
\end{figure}
Basis-size convergence of the polarization contributions to the EDM for $p$-shell nuclei is also quite reasonable and comparable to that of the anapole moments~\cite{Hao2020}. In Fig.~\ref{fig:LiBeconv}, we show the $N_{\rm max}$ convergence of the isovector $\pi$-exchange contribution for $^6$Li and $^9$Be as a representative example. Again, the the odd $N_{\rm max}$ values correspond to the unnatural-parity states in Eq.~(\ref{gswf}). The largest spaces that we were able to reach for $^{6,7}$Li were $N_{\rm max}{=}11$, while for $^9$Be $N_{\rm max}{=}9$. For $^{10,11}$B, our calculations have been performed up to $N_{\rm max}{=}7$. For $^{13}$C, $^{14,15}$N we also reached $N_{\rm max}{=}7$ basis space. However, we applied the importance truncation~\cite{PhysRevLett.99.092501,PhysRevC.79.064324} at $N_{\rm max}{=}7$ for these isotopes. The $^{19}$F is on the borderline of NCSM applicability. Only calculations up to $N_{\rm max}{=}5$ were performed although without any importance truncation. The $M$-scheme dimension was 189 million in this case.

\begin{figure*}
\centering
\includegraphics[width=\textwidth]{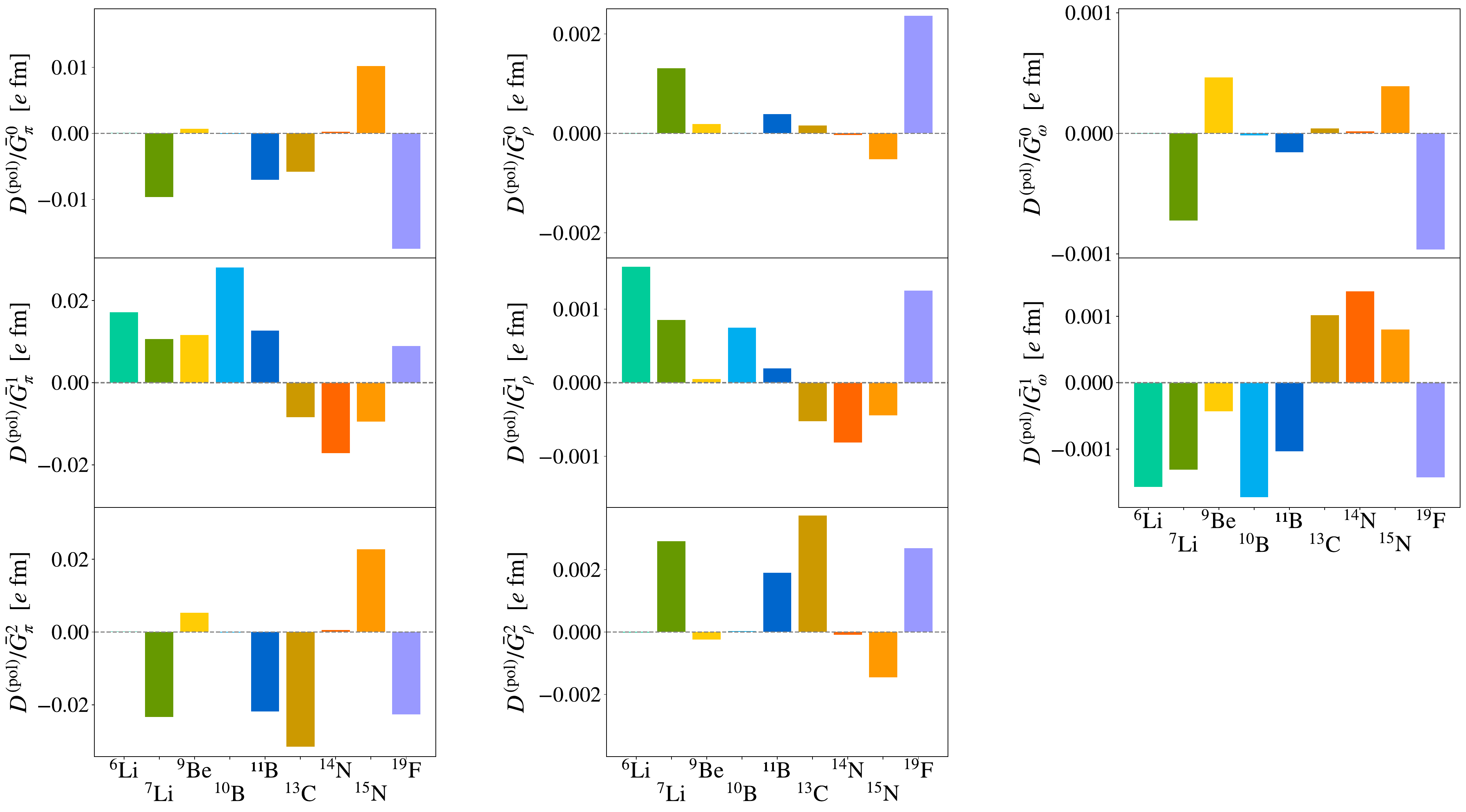}
\caption{\label{fig:Dpolsum} The polarization contribution to EDMs of stable $p$-shell nuclei and $^{19}$F (in $e$~fm) due to the $\chi$-exchange PTV NN interaction~(\ref{eq:PTVNN}), where $\chi$ stands for $\pi$, $\rho$, or $\omega$. SRG-evolved chiral NN+3N(lnl) PTC interaction from Ref.~\cite{PhysRevC.101.014318} was used.}
\end{figure*}
Our $D^{(1)}$ and $D^{({\rm pol})}$ results for all considered nuclei are shown in Table~\ref{tab1}. In Fig.~\ref{fig:Dpolsum}, we display all the calculated polarization contributions to the EDMs of the $p$-shell stable nuclei and $^{19}$F.  We can evaluate the uncertainties of our results due to the basis size convergence at about 10\% to 20\%. The other sources of uncertainty are renormalization and incompleteness of the transition operators and the uncertainties due to the description of the nuclear PTC and PTV forces. Although different sources of uncertainty might be at play, a rough estimate of the accuracy of our calculations can still be obtained by a comparison of the calculated and experimental magnetic moments shown in the last two columns of Table~\ref{tab1}. For $^{19}$F, we obtain in addition the magnetic moment +3.73~$\mu_{\rm N}$ for the $5/2^+$ excited state that can be compared to the experimental +3.607(8)~$\mu_{\rm N}$~\cite{STONE200575}. We note that we used a one-body M1 operator. The largest discrepancies occur for $^{11}$B and $^{13}$C from which we estimate the uncertainty of our results at about 30\%. 

The present results for $^{6,7}$Li, $^9$Be, $^{11}$B, and $^{13}$C nuclei can be compared to the cluster model calculations reported in Refs.~\cite{Yamanaka2015,Yamanaka2017a,Yamanaka2017b,Yamanaka2018,Yamanaka2019a,Yamanaka2019b}. For $^6$Li, cluster model results are available for $d_p$, $d_n$, and $\bar{G}^1_\chi$ contributions~\cite{Yamanaka2015,Yamanaka2017b,Yamanaka2018} and they are in a reasonable agreement with our calculations except for $\bar{G}^1_\omega$. For $^7$Li, available cluster model results for $d_p$ and $\bar{G}^T_\pi$~\cite{Yamanaka2018,Yamanaka2019b} are in a very good agreement with our {\it ab initio} calculations. For $^9$Be, our results for $d_n$ and $\bar{G}^1_\pi$ are close to those reported in Ref.~\cite{Yamanaka2018,Yamanaka2019a}. However, our  $\bar{G}^T_\omega$ results are smaller than the cluster model ones from Ref.~\cite{Yamanaka2015}. Our $^{11}$B results are within a factor of two of the cluster model calculations for $d_p$ and $\bar{G}^T_\pi$~\cite{Yamanaka2018,Yamanaka2019b}. For $^{13}$C, only $d_n$ and $\bar{G}^1_\pi$ cluster model results are avialable~\cite{Yamanaka2017a,Yamanaka2018}. While we are in agreement for the $d_n$, the {\it ab initio} NCSM result for the $\bar{G}^1_\pi$ contribution is larger by a factor of four. Interestingly, we get a significant isotensor $\bar{G}^2_\pi$ contribution that could not be calculated within the cluster model~\cite{Yamanaka2017a}.

As seen in Fig.~\ref{fig:Dpolsum}, our {\it ab initio} calculations show that different nuclei can be used to probe different terms of the parity violating interaction. For example, $^{10}$B has an enhanced $\bar{G}^1_\pi$ (by a factor of $\sim$2 compared to the deuteron~\cite{Liu2004}) as well as $\bar{G}^1_\omega$ contributions, $^6$Li the $\bar{G}^1_\rho$ contribution and $^{13}$C the $\bar{G}^2_\pi$ and $\bar{G}^2_\rho$. The $^{19}$F has dominant $D^{({\rm pol})}$contributions for several terms. This is to be expected to some extent as it has a low-lying $1/2^-$ state close to its $1/2^+$ ground state and overall high density of states compared to the $p$-shell nuclei. We also observe that the $D^{({\rm pol})}$ terms contribute by opposite signs for different nuclei.

\section{CONCLUSIONS AND OUTLOOK}

A nucleus in which a significantly enhanced $D^{({\rm pol})}$ can be anticipated is the exotic $^{11}$Be, famous for its ground-state parity inversion and the strongest known electric dipole transition between bound states~\cite{KWAN2014210}, with 13.8~s half-life that can be readily produced at facilities such as ISAC/ARIEL at TRIUMF. Due to the halo nature of its ground state, the NCSM used here is not applicable and rather the NCSM with continuum (NCSMC) must be used~\cite{Calci2016}. We are exploring a generalization of the present EDM calculation algorithms to NCSMC. 

The present calculations can be improved using the very recently developed chiral PTV interactions~\cite{deVries2020,Gnech2020,deVries21} instead of the one-meson-exchange model ones. The PTV NN interaction and the EDM operator should be SRG renormalized consistently with the nuclear chiral Hamiltonian. The technical capability to do this in the NCSM has been developed~\cite{Gysbers2019} and the renormalization calculations are under way. In general, the SRG transformation is mostly driven by short range correlations in the PTC NN interaction and its effect on longer-range operators such as the electric dipole, spin, and the leading order pion-exchange PTV interaction is expected to be rather small, i.e., a few percent~\cite{Gysbers2019,Schuster2014,Schuster2015,Miyagi2019}. The effect of the SRG transformation on short range parts of the PTV interaction due to the $\rho$- and $\omega$-exchange might be more significant and could reach $\sim$15\% (see, e.g., Fig. 3 in Ref.~\cite{Schuster2014} where a dependence on the operator range is discussed). Finally, two-body PTV operators could be included~\cite{Liu2004}.  

In summary, we performed {\it ab initio} calculations of EDMs of light nuclei beyond the typically studied $A{=}2,3$ systems. These calculations allow us to better understand which nuclei may have enhanced EDMs, and thus allow us to suggest which ones may be good candidates in the search for a measurable permanent electric dipole moment.

%

\acknowledgments

We thank I. Stetcu for useful discussions. This work was supported by the NSERC Grant No. SAPIN-2016-00033. TRIUMF receives federal funding via a contribution agreement with the National Research Council of Canada.  Computing support came from an INCITE Award on the Summit supercomputer of the Oak Ridge Leadership Computing Facility (OLCF) at ORNL, and from Westgrid and Compute Canada.


\begin{thebibliography}{58}%
\makeatletter
\providecommand \@ifxundefined [1]{%
 \@ifx{#1\undefined}
}%
\providecommand \@ifnum [1]{%
 \ifnum #1\expandafter \@firstoftwo
 \else \expandafter \@secondoftwo
 \fi
}%
\providecommand \@ifx [1]{%
 \ifx #1\expandafter \@firstoftwo
 \else \expandafter \@secondoftwo
 \fi
}%
\providecommand \natexlab [1]{#1}%
\providecommand \enquote  [1]{``#1''}%
\providecommand \bibnamefont  [1]{#1}%
\providecommand \bibfnamefont [1]{#1}%
\providecommand \citenamefont [1]{#1}%
\providecommand \href@noop [0]{\@secondoftwo}%
\providecommand \href [0]{\begingroup \@sanitize@url \@href}%
\providecommand \@href[1]{\@@startlink{#1}\@@href}%
\providecommand \@@href[1]{\endgroup#1\@@endlink}%
\providecommand \@sanitize@url [0]{\catcode `\\12\catcode `\$12\catcode
  `\&12\catcode `\#12\catcode `\^12\catcode `\_12\catcode `\%12\relax}%
\providecommand \@@startlink[1]{}%
\providecommand \@@endlink[0]{}%
\providecommand \url  [0]{\begingroup\@sanitize@url \@url }%
\providecommand \@url [1]{\endgroup\@href {#1}{\urlprefix }}%
\providecommand \urlprefix  [0]{URL }%
\providecommand \Eprint [0]{\href }%
\providecommand \doibase [0]{https://doi.org/}%
\providecommand \selectlanguage [0]{\@gobble}%
\providecommand \bibinfo  [0]{\@secondoftwo}%
\providecommand \bibfield  [0]{\@secondoftwo}%
\providecommand \translation [1]{[#1]}%
\providecommand \BibitemOpen [0]{}%
\providecommand \bibitemStop [0]{}%
\providecommand \bibitemNoStop [0]{.\EOS\space}%
\providecommand \EOS [0]{\spacefactor3000\relax}%
\providecommand \BibitemShut  [1]{\csname bibitem#1\endcsname}%
\let\auto@bib@innerbib\@empty
\bibitem [{\citenamefont {Sakharov}(1991)}]{Sakharov:1967dj}%
  \BibitemOpen
  \bibfield  {author} {\bibinfo {author} {\bibfnamefont {A.}~\bibnamefont
  {Sakharov}},\ }\bibfield  {title} {\bibinfo {title} {{Violation of CP
  Invariance, C asymmetry, and baryon asymmetry of the universe}},\ }\href
  {https://doi.org/10.1070/PU1991v034n05ABEH002497} {\bibfield  {journal}
  {\bibinfo  {journal} {Sov. Phys. Usp.}\ }\textbf {\bibinfo {volume} {34}},\
  \bibinfo {pages} {392} (\bibinfo {year} {1991})}\BibitemShut {NoStop}%
\bibitem [{\citenamefont {Kobayashi}\ and\ \citenamefont
  {Maskawa}(1973)}]{CKM1973}%
  \BibitemOpen
  \bibfield  {author} {\bibinfo {author} {\bibfnamefont {M.}~\bibnamefont
  {Kobayashi}}\ and\ \bibinfo {author} {\bibfnamefont {T.}~\bibnamefont
  {Maskawa}},\ }\bibfield  {title} {\bibinfo {title} {{CP-Violation in the
  Renormalizable Theory of Weak Interaction}},\ }\href
  {https://doi.org/10.1143/PTP.49.652} {\bibfield  {journal} {\bibinfo
  {journal} {Progress of Theoretical Physics}\ }\textbf {\bibinfo {volume}
  {49}},\ \bibinfo {pages} {652} (\bibinfo {year} {1973})},\ \Eprint
  {https://arxiv.org/abs/https://academic.oup.com/ptp/article-pdf/49/2/652/5257692/49-2-652.pdf}
  {https://academic.oup.com/ptp/article-pdf/49/2/652/5257692/49-2-652.pdf}
  \BibitemShut {NoStop}%
\bibitem [{\citenamefont {'t~Hooft}(1976)}]{PhysRevLett.37.8}%
  \BibitemOpen
  \bibfield  {author} {\bibinfo {author} {\bibfnamefont {G.}~\bibnamefont
  {'t~Hooft}},\ }\bibfield  {title} {\bibinfo {title} {Symmetry breaking
  through bell-jackiw anomalies},\ }\href
  {https://doi.org/10.1103/PhysRevLett.37.8} {\bibfield  {journal} {\bibinfo
  {journal} {Phys. Rev. Lett.}\ }\textbf {\bibinfo {volume} {37}},\ \bibinfo
  {pages} {8} (\bibinfo {year} {1976})}\BibitemShut {NoStop}%
\bibitem [{\citenamefont {Gavela}\ \emph
  {et~al.}(1994{\natexlab{a}})\citenamefont {Gavela}, \citenamefont {Lozano},
  \citenamefont {Orloff},\ and\ \citenamefont {Pene}}]{Gavela_1994a}%
  \BibitemOpen
  \bibfield  {author} {\bibinfo {author} {\bibfnamefont {M.}~\bibnamefont
  {Gavela}}, \bibinfo {author} {\bibfnamefont {M.}~\bibnamefont {Lozano}},
  \bibinfo {author} {\bibfnamefont {J.}~\bibnamefont {Orloff}},\ and\ \bibinfo
  {author} {\bibfnamefont {O.}~\bibnamefont {Pene}},\ }\bibfield  {title}
  {\bibinfo {title} {Standard model cp-violation and baryon asymmetry (i). zero
  temperature},\ }\href {https://doi.org/10.1016/0550-3213(94)00409-9}
  {\bibfield  {journal} {\bibinfo  {journal} {Nuclear Physics B}\ }\textbf
  {\bibinfo {volume} {430}},\ \bibinfo {pages} {345} (\bibinfo {year}
  {1994}{\natexlab{a}})}\BibitemShut {NoStop}%
\bibitem [{\citenamefont {Gavela}\ \emph
  {et~al.}(1994{\natexlab{b}})\citenamefont {Gavela}, \citenamefont
  {Hernandez}, \citenamefont {Orloff}, \citenamefont {Pene},\ and\
  \citenamefont {Quimbay}}]{Gavela_1994b}%
  \BibitemOpen
  \bibfield  {author} {\bibinfo {author} {\bibfnamefont {M.}~\bibnamefont
  {Gavela}}, \bibinfo {author} {\bibfnamefont {P.}~\bibnamefont {Hernandez}},
  \bibinfo {author} {\bibfnamefont {J.}~\bibnamefont {Orloff}}, \bibinfo
  {author} {\bibfnamefont {O.}~\bibnamefont {Pene}},\ and\ \bibinfo {author}
  {\bibfnamefont {C.}~\bibnamefont {Quimbay}},\ }\bibfield  {title} {\bibinfo
  {title} {Standard model cp-violation and baryon asymmetry (ii). finite
  temperature},\ }\href {https://doi.org/10.1016/0550-3213(94)00410-2}
  {\bibfield  {journal} {\bibinfo  {journal} {Nuclear Physics B}\ }\textbf
  {\bibinfo {volume} {430}},\ \bibinfo {pages} {382} (\bibinfo {year}
  {1994}{\natexlab{b}})}\BibitemShut {NoStop}%
\bibitem [{\citenamefont {Chupp}\ \emph {et~al.}(2019)\citenamefont {Chupp},
  \citenamefont {Fierlinger}, \citenamefont {Ramsey-Musolf},\ and\
  \citenamefont {Singh}}]{RevModPhys.91.015001}%
  \BibitemOpen
  \bibfield  {author} {\bibinfo {author} {\bibfnamefont {T.~E.}\ \bibnamefont
  {Chupp}}, \bibinfo {author} {\bibfnamefont {P.}~\bibnamefont {Fierlinger}},
  \bibinfo {author} {\bibfnamefont {M.~J.}\ \bibnamefont {Ramsey-Musolf}},\
  and\ \bibinfo {author} {\bibfnamefont {J.~T.}\ \bibnamefont {Singh}},\
  }\bibfield  {title} {\bibinfo {title} {Electric dipole moments of atoms,
  molecules, nuclei, and particles},\ }\href
  {https://doi.org/10.1103/RevModPhys.91.015001} {\bibfield  {journal}
  {\bibinfo  {journal} {Rev. Mod. Phys.}\ }\textbf {\bibinfo {volume} {91}},\
  \bibinfo {pages} {015001} (\bibinfo {year} {2019})}\BibitemShut {NoStop}%
\bibitem [{\citenamefont {Abel}\ \emph {et~al.}(2020)\citenamefont {Abel},
  \citenamefont {Afach}, \citenamefont {Ayres}, \citenamefont {Baker},
  \citenamefont {Ban}, \citenamefont {Bison}, \citenamefont {Bodek},
  \citenamefont {Bondar}, \citenamefont {Burghoff}, \citenamefont {Chanel},
  \citenamefont {Chowdhuri}, \citenamefont {Chiu}, \citenamefont {Clement},
  \citenamefont {Crawford}, \citenamefont {Daum}, \citenamefont {Emmenegger},
  \citenamefont {Ferraris-Bouchez}, \citenamefont {Fertl}, \citenamefont
  {Flaux}, \citenamefont {Franke}, \citenamefont {Fratangelo}, \citenamefont
  {Geltenbort}, \citenamefont {Green}, \citenamefont {Griffith}, \citenamefont
  {van~der Grinten}, \citenamefont {Gruji\ifmmode~\acute{c}\else \'{c}\fi{}},
  \citenamefont {Harris}, \citenamefont {Hayen}, \citenamefont {Heil},
  \citenamefont {Henneck}, \citenamefont {H\'elaine}, \citenamefont {Hild},
  \citenamefont {Hodge}, \citenamefont {Horras}, \citenamefont {Iaydjiev},
  \citenamefont {Ivanov}, \citenamefont {Kasprzak}, \citenamefont {Kermaidic},
  \citenamefont {Kirch}, \citenamefont {Knecht}, \citenamefont {Knowles},
  \citenamefont {Koch}, \citenamefont {Koss}, \citenamefont {Komposch},
  \citenamefont {Kozela}, \citenamefont {Kraft}, \citenamefont {Krempel},
  \citenamefont {Ku\ifmmode~\acute{z}\else \'{z}\fi{}niak}, \citenamefont
  {Lauss}, \citenamefont {Lefort}, \citenamefont {Lemi\`ere}, \citenamefont
  {Leredde}, \citenamefont {Mohanmurthy}, \citenamefont {Mtchedlishvili},
  \citenamefont {Musgrave}, \citenamefont {Naviliat-Cuncic}, \citenamefont
  {Pais}, \citenamefont {Piegsa}, \citenamefont {Pierre}, \citenamefont
  {Pignol}, \citenamefont {Plonka-Spehr}, \citenamefont {Prashanth},
  \citenamefont {Qu\'em\'ener}, \citenamefont {Rawlik}, \citenamefont
  {Rebreyend}, \citenamefont {Rien\"acker}, \citenamefont {Ries}, \citenamefont
  {Roccia}, \citenamefont {Rogel}, \citenamefont {Rozpedzik}, \citenamefont
  {Schnabel}, \citenamefont {Schmidt-Wellenburg}, \citenamefont {Severijns},
  \citenamefont {Shiers}, \citenamefont {Tavakoli~Dinani}, \citenamefont
  {Thorne}, \citenamefont {Virot}, \citenamefont {Voigt}, \citenamefont {Weis},
  \citenamefont {Wursten}, \citenamefont {Wyszynski}, \citenamefont {Zejma},
  \citenamefont {Zenner},\ and\ \citenamefont {Zsigmond}}]{nEDM2020}%
  \BibitemOpen
  \bibfield  {author} {\bibinfo {author} {\bibfnamefont {C.}~\bibnamefont
  {Abel}}, \bibinfo {author} {\bibfnamefont {S.}~\bibnamefont {Afach}},
  \bibinfo {author} {\bibfnamefont {N.~J.}\ \bibnamefont {Ayres}}, \bibinfo
  {author} {\bibfnamefont {C.~A.}\ \bibnamefont {Baker}}, \bibinfo {author}
  {\bibfnamefont {G.}~\bibnamefont {Ban}}, \bibinfo {author} {\bibfnamefont
  {G.}~\bibnamefont {Bison}}, \bibinfo {author} {\bibfnamefont
  {K.}~\bibnamefont {Bodek}}, \bibinfo {author} {\bibfnamefont
  {V.}~\bibnamefont {Bondar}}, \bibinfo {author} {\bibfnamefont
  {M.}~\bibnamefont {Burghoff}}, \bibinfo {author} {\bibfnamefont
  {E.}~\bibnamefont {Chanel}}, \bibinfo {author} {\bibfnamefont
  {Z.}~\bibnamefont {Chowdhuri}}, \bibinfo {author} {\bibfnamefont {P.-J.}\
  \bibnamefont {Chiu}}, \bibinfo {author} {\bibfnamefont {B.}~\bibnamefont
  {Clement}}, \bibinfo {author} {\bibfnamefont {C.~B.}\ \bibnamefont
  {Crawford}}, \bibinfo {author} {\bibfnamefont {M.}~\bibnamefont {Daum}},
  \bibinfo {author} {\bibfnamefont {S.}~\bibnamefont {Emmenegger}}, \bibinfo
  {author} {\bibfnamefont {L.}~\bibnamefont {Ferraris-Bouchez}}, \bibinfo
  {author} {\bibfnamefont {M.}~\bibnamefont {Fertl}}, \bibinfo {author}
  {\bibfnamefont {P.}~\bibnamefont {Flaux}}, \bibinfo {author} {\bibfnamefont
  {B.}~\bibnamefont {Franke}}, \bibinfo {author} {\bibfnamefont
  {A.}~\bibnamefont {Fratangelo}}, \bibinfo {author} {\bibfnamefont
  {P.}~\bibnamefont {Geltenbort}}, \bibinfo {author} {\bibfnamefont
  {K.}~\bibnamefont {Green}}, \bibinfo {author} {\bibfnamefont {W.~C.}\
  \bibnamefont {Griffith}}, \bibinfo {author} {\bibfnamefont {M.}~\bibnamefont
  {van~der Grinten}}, \bibinfo {author} {\bibfnamefont {Z.~D.}\ \bibnamefont
  {Gruji\ifmmode~\acute{c}\else \'{c}\fi{}}}, \bibinfo {author} {\bibfnamefont
  {P.~G.}\ \bibnamefont {Harris}}, \bibinfo {author} {\bibfnamefont
  {L.}~\bibnamefont {Hayen}}, \bibinfo {author} {\bibfnamefont
  {W.}~\bibnamefont {Heil}}, \bibinfo {author} {\bibfnamefont {R.}~\bibnamefont
  {Henneck}}, \bibinfo {author} {\bibfnamefont {V.}~\bibnamefont {H\'elaine}},
  \bibinfo {author} {\bibfnamefont {N.}~\bibnamefont {Hild}}, \bibinfo {author}
  {\bibfnamefont {Z.}~\bibnamefont {Hodge}}, \bibinfo {author} {\bibfnamefont
  {M.}~\bibnamefont {Horras}}, \bibinfo {author} {\bibfnamefont
  {P.}~\bibnamefont {Iaydjiev}}, \bibinfo {author} {\bibfnamefont {S.~N.}\
  \bibnamefont {Ivanov}}, \bibinfo {author} {\bibfnamefont {M.}~\bibnamefont
  {Kasprzak}}, \bibinfo {author} {\bibfnamefont {Y.}~\bibnamefont {Kermaidic}},
  \bibinfo {author} {\bibfnamefont {K.}~\bibnamefont {Kirch}}, \bibinfo
  {author} {\bibfnamefont {A.}~\bibnamefont {Knecht}}, \bibinfo {author}
  {\bibfnamefont {P.}~\bibnamefont {Knowles}}, \bibinfo {author} {\bibfnamefont
  {H.-C.}\ \bibnamefont {Koch}}, \bibinfo {author} {\bibfnamefont {P.~A.}\
  \bibnamefont {Koss}}, \bibinfo {author} {\bibfnamefont {S.}~\bibnamefont
  {Komposch}}, \bibinfo {author} {\bibfnamefont {A.}~\bibnamefont {Kozela}},
  \bibinfo {author} {\bibfnamefont {A.}~\bibnamefont {Kraft}}, \bibinfo
  {author} {\bibfnamefont {J.}~\bibnamefont {Krempel}}, \bibinfo {author}
  {\bibfnamefont {M.}~\bibnamefont {Ku\ifmmode~\acute{z}\else \'{z}\fi{}niak}},
  \bibinfo {author} {\bibfnamefont {B.}~\bibnamefont {Lauss}}, \bibinfo
  {author} {\bibfnamefont {T.}~\bibnamefont {Lefort}}, \bibinfo {author}
  {\bibfnamefont {Y.}~\bibnamefont {Lemi\`ere}}, \bibinfo {author}
  {\bibfnamefont {A.}~\bibnamefont {Leredde}}, \bibinfo {author} {\bibfnamefont
  {P.}~\bibnamefont {Mohanmurthy}}, \bibinfo {author} {\bibfnamefont
  {A.}~\bibnamefont {Mtchedlishvili}}, \bibinfo {author} {\bibfnamefont
  {M.}~\bibnamefont {Musgrave}}, \bibinfo {author} {\bibfnamefont
  {O.}~\bibnamefont {Naviliat-Cuncic}}, \bibinfo {author} {\bibfnamefont
  {D.}~\bibnamefont {Pais}}, \bibinfo {author} {\bibfnamefont {F.~M.}\
  \bibnamefont {Piegsa}}, \bibinfo {author} {\bibfnamefont {E.}~\bibnamefont
  {Pierre}}, \bibinfo {author} {\bibfnamefont {G.}~\bibnamefont {Pignol}},
  \bibinfo {author} {\bibfnamefont {C.}~\bibnamefont {Plonka-Spehr}}, \bibinfo
  {author} {\bibfnamefont {P.~N.}\ \bibnamefont {Prashanth}}, \bibinfo {author}
  {\bibfnamefont {G.}~\bibnamefont {Qu\'em\'ener}}, \bibinfo {author}
  {\bibfnamefont {M.}~\bibnamefont {Rawlik}}, \bibinfo {author} {\bibfnamefont
  {D.}~\bibnamefont {Rebreyend}}, \bibinfo {author} {\bibfnamefont
  {I.}~\bibnamefont {Rien\"acker}}, \bibinfo {author} {\bibfnamefont
  {D.}~\bibnamefont {Ries}}, \bibinfo {author} {\bibfnamefont {S.}~\bibnamefont
  {Roccia}}, \bibinfo {author} {\bibfnamefont {G.}~\bibnamefont {Rogel}},
  \bibinfo {author} {\bibfnamefont {D.}~\bibnamefont {Rozpedzik}}, \bibinfo
  {author} {\bibfnamefont {A.}~\bibnamefont {Schnabel}}, \bibinfo {author}
  {\bibfnamefont {P.}~\bibnamefont {Schmidt-Wellenburg}}, \bibinfo {author}
  {\bibfnamefont {N.}~\bibnamefont {Severijns}}, \bibinfo {author}
  {\bibfnamefont {D.}~\bibnamefont {Shiers}}, \bibinfo {author} {\bibfnamefont
  {R.}~\bibnamefont {Tavakoli~Dinani}}, \bibinfo {author} {\bibfnamefont
  {J.~A.}\ \bibnamefont {Thorne}}, \bibinfo {author} {\bibfnamefont
  {R.}~\bibnamefont {Virot}}, \bibinfo {author} {\bibfnamefont
  {J.}~\bibnamefont {Voigt}}, \bibinfo {author} {\bibfnamefont
  {A.}~\bibnamefont {Weis}}, \bibinfo {author} {\bibfnamefont {E.}~\bibnamefont
  {Wursten}}, \bibinfo {author} {\bibfnamefont {G.}~\bibnamefont {Wyszynski}},
  \bibinfo {author} {\bibfnamefont {J.}~\bibnamefont {Zejma}}, \bibinfo
  {author} {\bibfnamefont {J.}~\bibnamefont {Zenner}},\ and\ \bibinfo {author}
  {\bibfnamefont {G.}~\bibnamefont {Zsigmond}},\ }\bibfield  {title} {\bibinfo
  {title} {Measurement of the permanent electric dipole moment of the
  neutron},\ }\href {https://doi.org/10.1103/PhysRevLett.124.081803} {\bibfield
   {journal} {\bibinfo  {journal} {Phys. Rev. Lett.}\ }\textbf {\bibinfo
  {volume} {124}},\ \bibinfo {pages} {081803} (\bibinfo {year}
  {2020})}\BibitemShut {NoStop}%
\bibitem [{\citenamefont {Graner}\ \emph {et~al.}(2016)\citenamefont {Graner},
  \citenamefont {Chen}, \citenamefont {Lindahl},\ and\ \citenamefont
  {Heckel}}]{Graner2016}%
  \BibitemOpen
  \bibfield  {author} {\bibinfo {author} {\bibfnamefont {B.}~\bibnamefont
  {Graner}}, \bibinfo {author} {\bibfnamefont {Y.}~\bibnamefont {Chen}},
  \bibinfo {author} {\bibfnamefont {E.~G.}\ \bibnamefont {Lindahl}},\ and\
  \bibinfo {author} {\bibfnamefont {B.~R.}\ \bibnamefont {Heckel}},\ }\bibfield
   {title} {\bibinfo {title} {Reduced limit on the permanent electric dipole
  moment of $^{199}\mathrm{Hg}$},\ }\href
  {https://doi.org/10.1103/PhysRevLett.116.161601} {\bibfield  {journal}
  {\bibinfo  {journal} {Phys. Rev. Lett.}\ }\textbf {\bibinfo {volume} {116}},\
  \bibinfo {pages} {161601} (\bibinfo {year} {2016})}\BibitemShut {NoStop}%
\bibitem [{\citenamefont {Baron}\ \emph {et~al.}(2014)\citenamefont {Baron},
  \citenamefont {Campbell}, \citenamefont {DeMille}, \citenamefont {Doyle},
  \citenamefont {Gabrielse}, \citenamefont {Gurevich}, \citenamefont {Hess},
  \citenamefont {Hutzler}, \citenamefont {Kirilov}, \citenamefont {Kozyryev},
  \citenamefont {O{\textquoteright}Leary}, \citenamefont {Panda}, \citenamefont
  {Parsons}, \citenamefont {Petrik}, \citenamefont {Spaun}, \citenamefont
  {Vutha},\ and\ \citenamefont {West}}]{ACME2014}%
  \BibitemOpen
  \bibfield  {author} {\bibinfo {author} {\bibfnamefont {J.}~\bibnamefont
  {Baron}}, \bibinfo {author} {\bibfnamefont {W.~C.}\ \bibnamefont {Campbell}},
  \bibinfo {author} {\bibfnamefont {D.}~\bibnamefont {DeMille}}, \bibinfo
  {author} {\bibfnamefont {J.~M.}\ \bibnamefont {Doyle}}, \bibinfo {author}
  {\bibfnamefont {G.}~\bibnamefont {Gabrielse}}, \bibinfo {author}
  {\bibfnamefont {Y.~V.}\ \bibnamefont {Gurevich}}, \bibinfo {author}
  {\bibfnamefont {P.~W.}\ \bibnamefont {Hess}}, \bibinfo {author}
  {\bibfnamefont {N.~R.}\ \bibnamefont {Hutzler}}, \bibinfo {author}
  {\bibfnamefont {E.}~\bibnamefont {Kirilov}}, \bibinfo {author} {\bibfnamefont
  {I.}~\bibnamefont {Kozyryev}}, \bibinfo {author} {\bibfnamefont {B.~R.}\
  \bibnamefont {O{\textquoteright}Leary}}, \bibinfo {author} {\bibfnamefont
  {C.~D.}\ \bibnamefont {Panda}}, \bibinfo {author} {\bibfnamefont {M.~F.}\
  \bibnamefont {Parsons}}, \bibinfo {author} {\bibfnamefont {E.~S.}\
  \bibnamefont {Petrik}}, \bibinfo {author} {\bibfnamefont {B.}~\bibnamefont
  {Spaun}}, \bibinfo {author} {\bibfnamefont {A.~C.}\ \bibnamefont {Vutha}},\
  and\ \bibinfo {author} {\bibfnamefont {A.~D.}\ \bibnamefont {West}} (\bibinfo
  {collaboration} {ACME Collaboration}),\ }\bibfield  {title} {\bibinfo {title}
  {Order of magnitude smaller limit on the electric dipole moment of the
  electron},\ }\href {https://doi.org/10.1126/science.1248213} {\bibfield
  {journal} {\bibinfo  {journal} {Science}\ }\textbf {\bibinfo {volume}
  {343}},\ \bibinfo {pages} {269} (\bibinfo {year} {2014})}\BibitemShut
  {NoStop}%
\bibitem [{\citenamefont {Orlov}\ \emph {et~al.}(2006)\citenamefont {Orlov},
  \citenamefont {Morse},\ and\ \citenamefont {Semertzidis}}]{Orlov2006}%
  \BibitemOpen
  \bibfield  {author} {\bibinfo {author} {\bibfnamefont {Y.~F.}\ \bibnamefont
  {Orlov}}, \bibinfo {author} {\bibfnamefont {W.~M.}\ \bibnamefont {Morse}},\
  and\ \bibinfo {author} {\bibfnamefont {Y.~K.}\ \bibnamefont {Semertzidis}},\
  }\bibfield  {title} {\bibinfo {title} {Resonance method of
  electric-dipole-moment measurements in storage rings},\ }\href
  {https://doi.org/10.1103/PhysRevLett.96.214802} {\bibfield  {journal}
  {\bibinfo  {journal} {Phys. Rev. Lett.}\ }\textbf {\bibinfo {volume} {96}},\
  \bibinfo {pages} {214802} (\bibinfo {year} {2006})}\BibitemShut {NoStop}%
\bibitem [{\citenamefont {Pretz}(2013)}]{Pretz_2013}%
  \BibitemOpen
  \bibfield  {author} {\bibinfo {author} {\bibfnamefont {J.}~\bibnamefont
  {Pretz}},\ }\bibfield  {title} {\bibinfo {title} {Measurement of permanent
  electric dipole moments of charged hadrons in storage rings},\ }\href
  {https://doi.org/10.1007/s10751-013-0799-4} {\bibfield  {journal} {\bibinfo
  {journal} {Hyperfine Interactions}\ }\textbf {\bibinfo {volume} {214}},\
  \bibinfo {pages} {111} (\bibinfo {year} {2013})}\BibitemShut {NoStop}%
\bibitem [{\citenamefont {Hempelmann}\ \emph {et~al.}(2017)\citenamefont
  {Hempelmann}, \citenamefont {Hejny}, \citenamefont {Pretz}, \citenamefont
  {Stephenson}, \citenamefont {Augustyniak}, \citenamefont {Bagdasarian},
  \citenamefont {Bai}, \citenamefont {Barion}, \citenamefont {Berz},
  \citenamefont {Chekmenev}, \citenamefont {Ciullo}, \citenamefont {Dymov},
  \citenamefont {Etzkorn}, \citenamefont {Eversmann}, \citenamefont {Gaisser},
  \citenamefont {Gebel}, \citenamefont {Grigoryev}, \citenamefont {Grzonka},
  \citenamefont {Guidoboni}, \citenamefont {Hanraths}, \citenamefont
  {Heberling}, \citenamefont {Hetzel}, \citenamefont {Hinder}, \citenamefont
  {Kacharava}, \citenamefont {Kamerdzhiev}, \citenamefont {Keshelashvili},
  \citenamefont {Koop}, \citenamefont {Kulikov}, \citenamefont {Lehrach},
  \citenamefont {Lenisa}, \citenamefont {Lomidze}, \citenamefont {Lorentz},
  \citenamefont {Maanen}, \citenamefont {Macharashvili}, \citenamefont
  {Magiera}, \citenamefont {Mchedlishvili}, \citenamefont {Mey}, \citenamefont
  {M\"uller}, \citenamefont {Nass}, \citenamefont {Nikolaev}, \citenamefont
  {Pesce}, \citenamefont {Prasuhn}, \citenamefont {Rathmann}, \citenamefont
  {Rosenthal}, \citenamefont {Saleev}, \citenamefont {Schmidt}, \citenamefont
  {Semertzidis}, \citenamefont {Shmakova}, \citenamefont {Silenko},
  \citenamefont {Slim}, \citenamefont {Soltner}, \citenamefont {Stahl},
  \citenamefont {Stassen}, \citenamefont {Stockhorst}, \citenamefont
  {Str\"oher}, \citenamefont {Tabidze}, \citenamefont {Tagliente},
  \citenamefont {Talman}, \citenamefont {Th\"orngren~Engblom}, \citenamefont
  {Trinkel}, \citenamefont {Uzikov}, \citenamefont {Valdau}, \citenamefont
  {Valetov}, \citenamefont {Vassiliev}, \citenamefont {Weidemann},
  \citenamefont {Wro\'{n}ska}, \citenamefont {W\"ustner}, \citenamefont
  {Zupra\'{n}ski},\ and\ \citenamefont {Zurek}}]{Hempelmann2017}%
  \BibitemOpen
  \bibfield  {author} {\bibinfo {author} {\bibfnamefont {N.}~\bibnamefont
  {Hempelmann}}, \bibinfo {author} {\bibfnamefont {V.}~\bibnamefont {Hejny}},
  \bibinfo {author} {\bibfnamefont {J.}~\bibnamefont {Pretz}}, \bibinfo
  {author} {\bibfnamefont {E.}~\bibnamefont {Stephenson}}, \bibinfo {author}
  {\bibfnamefont {W.}~\bibnamefont {Augustyniak}}, \bibinfo {author}
  {\bibfnamefont {Z.}~\bibnamefont {Bagdasarian}}, \bibinfo {author}
  {\bibfnamefont {M.}~\bibnamefont {Bai}}, \bibinfo {author} {\bibfnamefont
  {L.}~\bibnamefont {Barion}}, \bibinfo {author} {\bibfnamefont
  {M.}~\bibnamefont {Berz}}, \bibinfo {author} {\bibfnamefont {S.}~\bibnamefont
  {Chekmenev}}, \bibinfo {author} {\bibfnamefont {G.}~\bibnamefont {Ciullo}},
  \bibinfo {author} {\bibfnamefont {S.}~\bibnamefont {Dymov}}, \bibinfo
  {author} {\bibfnamefont {F.-J.}\ \bibnamefont {Etzkorn}}, \bibinfo {author}
  {\bibfnamefont {D.}~\bibnamefont {Eversmann}}, \bibinfo {author}
  {\bibfnamefont {M.}~\bibnamefont {Gaisser}}, \bibinfo {author} {\bibfnamefont
  {R.}~\bibnamefont {Gebel}}, \bibinfo {author} {\bibfnamefont
  {K.}~\bibnamefont {Grigoryev}}, \bibinfo {author} {\bibfnamefont
  {D.}~\bibnamefont {Grzonka}}, \bibinfo {author} {\bibfnamefont
  {G.}~\bibnamefont {Guidoboni}}, \bibinfo {author} {\bibfnamefont
  {T.}~\bibnamefont {Hanraths}}, \bibinfo {author} {\bibfnamefont
  {D.}~\bibnamefont {Heberling}}, \bibinfo {author} {\bibfnamefont
  {J.}~\bibnamefont {Hetzel}}, \bibinfo {author} {\bibfnamefont
  {F.}~\bibnamefont {Hinder}}, \bibinfo {author} {\bibfnamefont
  {A.}~\bibnamefont {Kacharava}}, \bibinfo {author} {\bibfnamefont
  {V.}~\bibnamefont {Kamerdzhiev}}, \bibinfo {author} {\bibfnamefont
  {I.}~\bibnamefont {Keshelashvili}}, \bibinfo {author} {\bibfnamefont
  {I.}~\bibnamefont {Koop}}, \bibinfo {author} {\bibfnamefont {A.}~\bibnamefont
  {Kulikov}}, \bibinfo {author} {\bibfnamefont {A.}~\bibnamefont {Lehrach}},
  \bibinfo {author} {\bibfnamefont {P.}~\bibnamefont {Lenisa}}, \bibinfo
  {author} {\bibfnamefont {N.}~\bibnamefont {Lomidze}}, \bibinfo {author}
  {\bibfnamefont {B.}~\bibnamefont {Lorentz}}, \bibinfo {author} {\bibfnamefont
  {P.}~\bibnamefont {Maanen}}, \bibinfo {author} {\bibfnamefont
  {G.}~\bibnamefont {Macharashvili}}, \bibinfo {author} {\bibfnamefont
  {A.}~\bibnamefont {Magiera}}, \bibinfo {author} {\bibfnamefont
  {D.}~\bibnamefont {Mchedlishvili}}, \bibinfo {author} {\bibfnamefont
  {S.}~\bibnamefont {Mey}}, \bibinfo {author} {\bibfnamefont {F.}~\bibnamefont
  {M\"uller}}, \bibinfo {author} {\bibfnamefont {A.}~\bibnamefont {Nass}},
  \bibinfo {author} {\bibfnamefont {N.~N.}\ \bibnamefont {Nikolaev}}, \bibinfo
  {author} {\bibfnamefont {A.}~\bibnamefont {Pesce}}, \bibinfo {author}
  {\bibfnamefont {D.}~\bibnamefont {Prasuhn}}, \bibinfo {author} {\bibfnamefont
  {F.}~\bibnamefont {Rathmann}}, \bibinfo {author} {\bibfnamefont
  {M.}~\bibnamefont {Rosenthal}}, \bibinfo {author} {\bibfnamefont
  {A.}~\bibnamefont {Saleev}}, \bibinfo {author} {\bibfnamefont
  {V.}~\bibnamefont {Schmidt}}, \bibinfo {author} {\bibfnamefont
  {Y.}~\bibnamefont {Semertzidis}}, \bibinfo {author} {\bibfnamefont
  {V.}~\bibnamefont {Shmakova}}, \bibinfo {author} {\bibfnamefont
  {A.}~\bibnamefont {Silenko}}, \bibinfo {author} {\bibfnamefont
  {J.}~\bibnamefont {Slim}}, \bibinfo {author} {\bibfnamefont {H.}~\bibnamefont
  {Soltner}}, \bibinfo {author} {\bibfnamefont {A.}~\bibnamefont {Stahl}},
  \bibinfo {author} {\bibfnamefont {R.}~\bibnamefont {Stassen}}, \bibinfo
  {author} {\bibfnamefont {H.}~\bibnamefont {Stockhorst}}, \bibinfo {author}
  {\bibfnamefont {H.}~\bibnamefont {Str\"oher}}, \bibinfo {author}
  {\bibfnamefont {M.}~\bibnamefont {Tabidze}}, \bibinfo {author} {\bibfnamefont
  {G.}~\bibnamefont {Tagliente}}, \bibinfo {author} {\bibfnamefont
  {R.}~\bibnamefont {Talman}}, \bibinfo {author} {\bibfnamefont
  {P.}~\bibnamefont {Th\"orngren~Engblom}}, \bibinfo {author} {\bibfnamefont
  {F.}~\bibnamefont {Trinkel}}, \bibinfo {author} {\bibfnamefont
  {Y.}~\bibnamefont {Uzikov}}, \bibinfo {author} {\bibfnamefont
  {Y.}~\bibnamefont {Valdau}}, \bibinfo {author} {\bibfnamefont
  {E.}~\bibnamefont {Valetov}}, \bibinfo {author} {\bibfnamefont
  {A.}~\bibnamefont {Vassiliev}}, \bibinfo {author} {\bibfnamefont
  {C.}~\bibnamefont {Weidemann}}, \bibinfo {author} {\bibfnamefont
  {A.}~\bibnamefont {Wro\'{n}ska}}, \bibinfo {author} {\bibfnamefont
  {P.}~\bibnamefont {W\"ustner}}, \bibinfo {author} {\bibfnamefont
  {P.}~\bibnamefont {Zupra\'{n}ski}},\ and\ \bibinfo {author} {\bibfnamefont
  {M.}~\bibnamefont {Zurek}} (\bibinfo {collaboration} {JEDI Collaboration}),\
  }\bibfield  {title} {\bibinfo {title} {Phase locking the spin precession in a
  storage ring},\ }\href {https://doi.org/10.1103/PhysRevLett.119.014801}
  {\bibfield  {journal} {\bibinfo  {journal} {Phys. Rev. Lett.}\ }\textbf
  {\bibinfo {volume} {119}},\ \bibinfo {pages} {014801} (\bibinfo {year}
  {2017})}\BibitemShut {NoStop}%
\bibitem [{\citenamefont {Abusaif}\ \emph {et~al.}(2019)\citenamefont
  {Abusaif}, \citenamefont {Aggarwal}, \citenamefont {Aksentev}, \citenamefont
  {Alberdi-Esuain}, \citenamefont {Atanasov}, \citenamefont {Barion},
  \citenamefont {Basile}, \citenamefont {Berz}, \citenamefont {Beyss},
  \citenamefont {B\"{o}hme}, \citenamefont {B\"{o}ker}, \citenamefont
  {Borburgh}, \citenamefont {Carli}, \citenamefont {Ciepal}, \citenamefont
  {Ciullo}, \citenamefont {Contalbrigo}, \citenamefont {Conto}, \citenamefont
  {Dymov}, \citenamefont {Felden}, \citenamefont {Gagoshidze}, \citenamefont
  {Gaisser}, \citenamefont {Gebel}, \citenamefont {Giese}, \citenamefont
  {Grigoryev}, \citenamefont {Grzonka}, \citenamefont {Tahar}, \citenamefont
  {Hahnraths}, \citenamefont {Heberling}, \citenamefont {Hejny}, \citenamefont
  {Hetzel}, \citenamefont {H\"{o}lscher}, \citenamefont {Javakhishvili},
  \citenamefont {Jorat}, \citenamefont {Kacharava}, \citenamefont
  {Kamerdzhiev}, \citenamefont {Karanth}, \citenamefont {K\"{a}seberg},
  \citenamefont {Keshelashvili}, \citenamefont {Koop}, \citenamefont {Kulikov},
  \citenamefont {Laihem}, \citenamefont {Lamont}, \citenamefont {Lehrach},
  \citenamefont {Lenisa}, \citenamefont {Lomidze}, \citenamefont {Lorentz},
  \citenamefont {Macharashvili}, \citenamefont {Magiera}, \citenamefont
  {Makino}, \citenamefont {Martin}, \citenamefont {Mchedlishvili},
  \citenamefont {Meissner}, \citenamefont {Metreveli}, \citenamefont {Michaud},
  \citenamefont {M\"{u}ller}, \citenamefont {Nass}, \citenamefont {Natour},
  \citenamefont {Nikolaev}, \citenamefont {Nogga}, \citenamefont {Pesce},
  \citenamefont {Poncza}, \citenamefont {Prasuhn}, \citenamefont {Pretz},
  \citenamefont {Rathmann}, \citenamefont {Ritman}, \citenamefont {Rosenthal},
  \citenamefont {Saleev}, \citenamefont {Schott}, \citenamefont {Sefzick},
  \citenamefont {Senichev}, \citenamefont {Shergelashvili}, \citenamefont
  {Shmakova}, \citenamefont {Siddique}, \citenamefont {Silenko}, \citenamefont
  {Simon}, \citenamefont {Slim}, \citenamefont {Soltner}, \citenamefont
  {Stahl}, \citenamefont {Stassen}, \citenamefont {Stephenson}, \citenamefont
  {Straatmann}, \citenamefont {Str\"{o}her}, \citenamefont {Tabidze},
  \citenamefont {Tagliente}, \citenamefont {Talman}, \citenamefont {Uzikov},
  \citenamefont {Valdau}, \citenamefont {Valetov}, \citenamefont {Wagner},
  \citenamefont {Weidemann}, \citenamefont {Wirzba}, \citenamefont
  {Wro\'{n}ska}, \citenamefont {W\"{u}stner}, \citenamefont {Zupra\'nski},\
  and\ \citenamefont {Zurek}}]{abusaif2019storage}%
  \BibitemOpen
  \bibfield  {author} {\bibinfo {author} {\bibfnamefont {F.}~\bibnamefont
  {Abusaif}}, \bibinfo {author} {\bibfnamefont {A.}~\bibnamefont {Aggarwal}},
  \bibinfo {author} {\bibfnamefont {A.}~\bibnamefont {Aksentev}}, \bibinfo
  {author} {\bibfnamefont {B.}~\bibnamefont {Alberdi-Esuain}}, \bibinfo
  {author} {\bibfnamefont {A.}~\bibnamefont {Atanasov}}, \bibinfo {author}
  {\bibfnamefont {L.}~\bibnamefont {Barion}}, \bibinfo {author} {\bibfnamefont
  {S.}~\bibnamefont {Basile}}, \bibinfo {author} {\bibfnamefont
  {M.}~\bibnamefont {Berz}}, \bibinfo {author} {\bibfnamefont {M.}~\bibnamefont
  {Beyss}}, \bibinfo {author} {\bibfnamefont {C.}~\bibnamefont {B\"{o}hme}},
  \bibinfo {author} {\bibfnamefont {J.}~\bibnamefont {B\"{o}ker}}, \bibinfo
  {author} {\bibfnamefont {J.}~\bibnamefont {Borburgh}}, \bibinfo {author}
  {\bibfnamefont {C.}~\bibnamefont {Carli}}, \bibinfo {author} {\bibfnamefont
  {I.}~\bibnamefont {Ciepal}}, \bibinfo {author} {\bibfnamefont
  {G.}~\bibnamefont {Ciullo}}, \bibinfo {author} {\bibfnamefont
  {M.}~\bibnamefont {Contalbrigo}}, \bibinfo {author} {\bibfnamefont
  {J.~M.~D.}\ \bibnamefont {Conto}}, \bibinfo {author} {\bibfnamefont
  {S.}~\bibnamefont {Dymov}}, \bibinfo {author} {\bibfnamefont
  {O.}~\bibnamefont {Felden}}, \bibinfo {author} {\bibfnamefont
  {M.}~\bibnamefont {Gagoshidze}}, \bibinfo {author} {\bibfnamefont
  {M.}~\bibnamefont {Gaisser}}, \bibinfo {author} {\bibfnamefont
  {R.}~\bibnamefont {Gebel}}, \bibinfo {author} {\bibfnamefont
  {N.}~\bibnamefont {Giese}}, \bibinfo {author} {\bibfnamefont
  {K.}~\bibnamefont {Grigoryev}}, \bibinfo {author} {\bibfnamefont
  {D.}~\bibnamefont {Grzonka}}, \bibinfo {author} {\bibfnamefont {M.~H.}\
  \bibnamefont {Tahar}}, \bibinfo {author} {\bibfnamefont {T.}~\bibnamefont
  {Hahnraths}}, \bibinfo {author} {\bibfnamefont {D.}~\bibnamefont
  {Heberling}}, \bibinfo {author} {\bibfnamefont {V.}~\bibnamefont {Hejny}},
  \bibinfo {author} {\bibfnamefont {J.}~\bibnamefont {Hetzel}}, \bibinfo
  {author} {\bibfnamefont {D.}~\bibnamefont {H\"{o}lscher}}, \bibinfo {author}
  {\bibfnamefont {O.}~\bibnamefont {Javakhishvili}}, \bibinfo {author}
  {\bibfnamefont {L.}~\bibnamefont {Jorat}}, \bibinfo {author} {\bibfnamefont
  {A.}~\bibnamefont {Kacharava}}, \bibinfo {author} {\bibfnamefont
  {V.}~\bibnamefont {Kamerdzhiev}}, \bibinfo {author} {\bibfnamefont
  {S.}~\bibnamefont {Karanth}}, \bibinfo {author} {\bibfnamefont
  {C.}~\bibnamefont {K\"{a}seberg}}, \bibinfo {author} {\bibfnamefont
  {I.}~\bibnamefont {Keshelashvili}}, \bibinfo {author} {\bibfnamefont
  {I.}~\bibnamefont {Koop}}, \bibinfo {author} {\bibfnamefont {A.}~\bibnamefont
  {Kulikov}}, \bibinfo {author} {\bibfnamefont {K.}~\bibnamefont {Laihem}},
  \bibinfo {author} {\bibfnamefont {M.}~\bibnamefont {Lamont}}, \bibinfo
  {author} {\bibfnamefont {A.}~\bibnamefont {Lehrach}}, \bibinfo {author}
  {\bibfnamefont {P.}~\bibnamefont {Lenisa}}, \bibinfo {author} {\bibfnamefont
  {N.}~\bibnamefont {Lomidze}}, \bibinfo {author} {\bibfnamefont
  {B.}~\bibnamefont {Lorentz}}, \bibinfo {author} {\bibfnamefont
  {G.}~\bibnamefont {Macharashvili}}, \bibinfo {author} {\bibfnamefont
  {A.}~\bibnamefont {Magiera}}, \bibinfo {author} {\bibfnamefont
  {K.}~\bibnamefont {Makino}}, \bibinfo {author} {\bibfnamefont
  {S.}~\bibnamefont {Martin}}, \bibinfo {author} {\bibfnamefont
  {D.}~\bibnamefont {Mchedlishvili}}, \bibinfo {author} {\bibfnamefont {U.~G.}\
  \bibnamefont {Meissner}}, \bibinfo {author} {\bibfnamefont {Z.}~\bibnamefont
  {Metreveli}}, \bibinfo {author} {\bibfnamefont {J.}~\bibnamefont {Michaud}},
  \bibinfo {author} {\bibfnamefont {F.}~\bibnamefont {M\"{u}ller}}, \bibinfo
  {author} {\bibfnamefont {A.}~\bibnamefont {Nass}}, \bibinfo {author}
  {\bibfnamefont {G.}~\bibnamefont {Natour}}, \bibinfo {author} {\bibfnamefont
  {N.}~\bibnamefont {Nikolaev}}, \bibinfo {author} {\bibfnamefont
  {A.}~\bibnamefont {Nogga}}, \bibinfo {author} {\bibfnamefont
  {A.}~\bibnamefont {Pesce}}, \bibinfo {author} {\bibfnamefont
  {V.}~\bibnamefont {Poncza}}, \bibinfo {author} {\bibfnamefont
  {D.}~\bibnamefont {Prasuhn}}, \bibinfo {author} {\bibfnamefont
  {J.}~\bibnamefont {Pretz}}, \bibinfo {author} {\bibfnamefont
  {F.}~\bibnamefont {Rathmann}}, \bibinfo {author} {\bibfnamefont
  {J.}~\bibnamefont {Ritman}}, \bibinfo {author} {\bibfnamefont
  {M.}~\bibnamefont {Rosenthal}}, \bibinfo {author} {\bibfnamefont
  {A.}~\bibnamefont {Saleev}}, \bibinfo {author} {\bibfnamefont
  {M.}~\bibnamefont {Schott}}, \bibinfo {author} {\bibfnamefont
  {T.}~\bibnamefont {Sefzick}}, \bibinfo {author} {\bibfnamefont
  {Y.}~\bibnamefont {Senichev}}, \bibinfo {author} {\bibfnamefont
  {D.}~\bibnamefont {Shergelashvili}}, \bibinfo {author} {\bibfnamefont
  {V.}~\bibnamefont {Shmakova}}, \bibinfo {author} {\bibfnamefont
  {S.}~\bibnamefont {Siddique}}, \bibinfo {author} {\bibfnamefont
  {A.}~\bibnamefont {Silenko}}, \bibinfo {author} {\bibfnamefont
  {M.}~\bibnamefont {Simon}}, \bibinfo {author} {\bibfnamefont
  {J.}~\bibnamefont {Slim}}, \bibinfo {author} {\bibfnamefont {H.}~\bibnamefont
  {Soltner}}, \bibinfo {author} {\bibfnamefont {A.}~\bibnamefont {Stahl}},
  \bibinfo {author} {\bibfnamefont {R.}~\bibnamefont {Stassen}}, \bibinfo
  {author} {\bibfnamefont {E.}~\bibnamefont {Stephenson}}, \bibinfo {author}
  {\bibfnamefont {H.}~\bibnamefont {Straatmann}}, \bibinfo {author}
  {\bibfnamefont {H.}~\bibnamefont {Str\"{o}her}}, \bibinfo {author}
  {\bibfnamefont {M.}~\bibnamefont {Tabidze}}, \bibinfo {author} {\bibfnamefont
  {G.}~\bibnamefont {Tagliente}}, \bibinfo {author} {\bibfnamefont
  {R.}~\bibnamefont {Talman}}, \bibinfo {author} {\bibfnamefont
  {Y.}~\bibnamefont {Uzikov}}, \bibinfo {author} {\bibfnamefont
  {Y.}~\bibnamefont {Valdau}}, \bibinfo {author} {\bibfnamefont
  {E.}~\bibnamefont {Valetov}}, \bibinfo {author} {\bibfnamefont
  {T.}~\bibnamefont {Wagner}}, \bibinfo {author} {\bibfnamefont
  {C.}~\bibnamefont {Weidemann}}, \bibinfo {author} {\bibfnamefont
  {A.}~\bibnamefont {Wirzba}}, \bibinfo {author} {\bibfnamefont
  {A.}~\bibnamefont {Wro\'{n}ska}}, \bibinfo {author} {\bibfnamefont
  {P.}~\bibnamefont {W\"{u}stner}}, \bibinfo {author} {\bibfnamefont
  {P.}~\bibnamefont {Zupra\'nski}},\ and\ \bibinfo {author} {\bibfnamefont
  {M.}~\bibnamefont {Zurek}},\ }\href@noop {} {\bibinfo {title} {Storage ring
  to search for electric dipole moments of charged particles: feasibility
  study}} (\bibinfo {year} {2019}),\ \Eprint {https://arxiv.org/abs/1912.07881}
{arXiv:1912.07881 [hep-ex]}
\href{https://doi.org/10.23731/CYRM-2021-003}{https://doi.org/10.23731/CYRM-2021-003}
\BibitemShut {NoStop}%
\bibitem [{\citenamefont {Schiff}(1963)}]{Schiff1963}%
  \BibitemOpen
  \bibfield  {author} {\bibinfo {author} {\bibfnamefont {L.~I.}\ \bibnamefont
  {Schiff}},\ }\bibfield  {title} {\bibinfo {title} {Measurability of nuclear
  electric dipole moments},\ }\href {https://doi.org/10.1103/PhysRev.132.2194}
  {\bibfield  {journal} {\bibinfo  {journal} {Phys. Rev.}\ }\textbf {\bibinfo
  {volume} {132}},\ \bibinfo {pages} {2194} (\bibinfo {year}
  {1963})}\BibitemShut {NoStop}%
\bibitem [{\citenamefont {Flambaum}\ \emph {et~al.}(1985)\citenamefont
  {Flambaum}, \citenamefont {Khriplovich},\ and\ \citenamefont
  {Sushkov}}]{Flambaum1985}%
  \BibitemOpen
  \bibfield  {author} {\bibinfo {author} {\bibfnamefont {V.}~\bibnamefont
  {Flambaum}}, \bibinfo {author} {\bibfnamefont {I.}~\bibnamefont
  {Khriplovich}},\ and\ \bibinfo {author} {\bibfnamefont {O.}~\bibnamefont
  {Sushkov}},\ }\bibfield  {title} {\bibinfo {title} {Limit on the constant of
  t-nonconserving nucleon-nucleon interaction},\ }\href
  {https://doi.org/https://doi.org/10.1016/0370-2693(85)90908-6} {\bibfield
  {journal} {\bibinfo  {journal} {Physics Letters B}\ }\textbf {\bibinfo
  {volume} {162}},\ \bibinfo {pages} {213 } (\bibinfo {year}
  {1985})}\BibitemShut {NoStop}%
\bibitem [{\citenamefont {Liu}\ and\ \citenamefont
  {Timmermans}(2004)}]{Liu2004}%
  \BibitemOpen
  \bibfield  {author} {\bibinfo {author} {\bibfnamefont {C.-P.}\ \bibnamefont
  {Liu}}\ and\ \bibinfo {author} {\bibfnamefont {R.~G.~E.}\ \bibnamefont
  {Timmermans}},\ }\bibfield  {title} {\bibinfo {title} {$p$- and $t$-odd
  two-nucleon interaction and the deuteron electric dipole moment},\ }\href
  {https://doi.org/10.1103/PhysRevC.70.055501} {\bibfield  {journal} {\bibinfo
  {journal} {Phys. Rev. C}\ }\textbf {\bibinfo {volume} {70}},\ \bibinfo
  {pages} {055501} (\bibinfo {year} {2004})}\BibitemShut {NoStop}%
\bibitem [{\citenamefont {Stetcu}\ \emph {et~al.}(2008)\citenamefont {Stetcu},
  \citenamefont {Liu}, \citenamefont {Friar}, \citenamefont {Hayes},\ and\
  \citenamefont {Navr\'atil}}]{STETCU2008168}%
  \BibitemOpen
  \bibfield  {author} {\bibinfo {author} {\bibfnamefont {I.}~\bibnamefont
  {Stetcu}}, \bibinfo {author} {\bibfnamefont {C.-P.}\ \bibnamefont {Liu}},
  \bibinfo {author} {\bibfnamefont {J.}~\bibnamefont {Friar}}, \bibinfo
  {author} {\bibfnamefont {A.}~\bibnamefont {Hayes}},\ and\ \bibinfo {author}
  {\bibfnamefont {P.}~\bibnamefont {Navr\'atil}},\ }\bibfield  {title}
  {\bibinfo {title} {Nuclear electric dipole moment of 3he},\ }\href
  {https://doi.org/https://doi.org/10.1016/j.physletb.2008.06.019} {\bibfield
  {journal} {\bibinfo  {journal} {Physics Letters B}\ }\textbf {\bibinfo
  {volume} {665}},\ \bibinfo {pages} {168 } (\bibinfo {year}
  {2008})}\BibitemShut {NoStop}%
\bibitem [{\citenamefont {de~Vries}\ \emph
  {et~al.}(2011{\natexlab{a}})\citenamefont {de~Vries}, \citenamefont
  {Mereghetti}, \citenamefont {Timmermans},\ and\ \citenamefont {van
  Kolck}}]{deVries2011d}%
  \BibitemOpen
  \bibfield  {author} {\bibinfo {author} {\bibfnamefont {J.}~\bibnamefont
  {de~Vries}}, \bibinfo {author} {\bibfnamefont {E.}~\bibnamefont
  {Mereghetti}}, \bibinfo {author} {\bibfnamefont {R.~G.~E.}\ \bibnamefont
  {Timmermans}},\ and\ \bibinfo {author} {\bibfnamefont {U.}~\bibnamefont {van
  Kolck}},\ }\bibfield  {title} {\bibinfo {title} {$p$ and $t$ violating form
  factors of the deuteron},\ }\href
  {https://doi.org/10.1103/PhysRevLett.107.091804} {\bibfield  {journal}
  {\bibinfo  {journal} {Phys. Rev. Lett.}\ }\textbf {\bibinfo {volume} {107}},\
  \bibinfo {pages} {091804} (\bibinfo {year} {2011}{\natexlab{a}})}\BibitemShut
  {NoStop}%
\bibitem [{\citenamefont {de~Vries}\ \emph
  {et~al.}(2011{\natexlab{b}})\citenamefont {de~Vries}, \citenamefont {Higa},
  \citenamefont {Liu}, \citenamefont {Mereghetti}, \citenamefont {Stetcu},
  \citenamefont {Timmermans},\ and\ \citenamefont {van Kolck}}]{deVries2011}%
  \BibitemOpen
  \bibfield  {author} {\bibinfo {author} {\bibfnamefont {J.}~\bibnamefont
  {de~Vries}}, \bibinfo {author} {\bibfnamefont {R.}~\bibnamefont {Higa}},
  \bibinfo {author} {\bibfnamefont {C.-P.}\ \bibnamefont {Liu}}, \bibinfo
  {author} {\bibfnamefont {E.}~\bibnamefont {Mereghetti}}, \bibinfo {author}
  {\bibfnamefont {I.}~\bibnamefont {Stetcu}}, \bibinfo {author} {\bibfnamefont
  {R.~G.~E.}\ \bibnamefont {Timmermans}},\ and\ \bibinfo {author}
  {\bibfnamefont {U.}~\bibnamefont {van Kolck}},\ }\bibfield  {title} {\bibinfo
  {title} {Electric dipole moments of light nuclei from chiral effective field
  theory},\ }\href {https://doi.org/10.1103/PhysRevC.84.065501} {\bibfield
  {journal} {\bibinfo  {journal} {Phys. Rev. C}\ }\textbf {\bibinfo {volume}
  {84}},\ \bibinfo {pages} {065501} (\bibinfo {year}
  {2011}{\natexlab{b}})}\BibitemShut {NoStop}%
\bibitem [{\citenamefont {Song}\ \emph {et~al.}(2013)\citenamefont {Song},
  \citenamefont {Lazauskas},\ and\ \citenamefont {Gudkov}}]{Lazauskas2013}%
  \BibitemOpen
  \bibfield  {author} {\bibinfo {author} {\bibfnamefont {Y.-H.}\ \bibnamefont
  {Song}}, \bibinfo {author} {\bibfnamefont {R.}~\bibnamefont {Lazauskas}},\
  and\ \bibinfo {author} {\bibfnamefont {V.}~\bibnamefont {Gudkov}},\
  }\bibfield  {title} {\bibinfo {title} {Nuclear electric dipole moment of
  three-body systems},\ }\href {https://doi.org/10.1103/PhysRevC.87.015501}
  {\bibfield  {journal} {\bibinfo  {journal} {Phys. Rev. C}\ }\textbf {\bibinfo
  {volume} {87}},\ \bibinfo {pages} {015501} (\bibinfo {year}
  {2013})}\BibitemShut {NoStop}%
\bibitem [{\citenamefont {Bsaisou}\ \emph {et~al.}(2015)\citenamefont
  {Bsaisou}, \citenamefont {Meissner}, \citenamefont {Nogga},\ and\
  \citenamefont {Wirzba}}]{Bsaisou2015}%
  \BibitemOpen
  \bibfield  {author} {\bibinfo {author} {\bibfnamefont {J.}~\bibnamefont
  {Bsaisou}}, \bibinfo {author} {\bibfnamefont {U.-G.}\ \bibnamefont
  {Meissner}}, \bibinfo {author} {\bibfnamefont {A.}~\bibnamefont {Nogga}},\
  and\ \bibinfo {author} {\bibfnamefont {A.}~\bibnamefont {Wirzba}},\
  }\bibfield  {title} {\bibinfo {title} {P- and t-violating lagrangians in
  chiral effective field theory and nuclear electric dipole moments},\ }\href
  {https://doi.org/https://doi.org/10.1016/j.aop.2015.04.031} {\bibfield
  {journal} {\bibinfo  {journal} {Annals of Physics}\ }\textbf {\bibinfo
  {volume} {359}},\ \bibinfo {pages} {317 } (\bibinfo {year}
  {2015})}\BibitemShut {NoStop}%
\bibitem [{\citenamefont {Wirzba}\ \emph {et~al.}(2017)\citenamefont {Wirzba},
  \citenamefont {Bsaisou},\ and\ \citenamefont {Nogga}}]{Wirzba2017}%
  \BibitemOpen
  \bibfield  {author} {\bibinfo {author} {\bibfnamefont {A.}~\bibnamefont
  {Wirzba}}, \bibinfo {author} {\bibfnamefont {J.}~\bibnamefont {Bsaisou}},\
  and\ \bibinfo {author} {\bibfnamefont {A.}~\bibnamefont {Nogga}},\ }\bibfield
   {title} {\bibinfo {title} {Permanent electric dipole moments of single-,
  two- and three-nucleon systems},\ }\href
  {https://doi.org/10.1142/S0218301317400316} {\bibfield  {journal} {\bibinfo
  {journal} {International Journal of Modern Physics E}\ }\textbf {\bibinfo
  {volume} {26}},\ \bibinfo {pages} {1740031} (\bibinfo {year} {2017})},\
  \Eprint {https://arxiv.org/abs/https://doi.org/10.1142/S0218301317400316}
  {https://doi.org/10.1142/S0218301317400316} \BibitemShut {NoStop}%
\bibitem [{\citenamefont {Gnech}\ and\ \citenamefont
  {Viviani}(2020)}]{Gnech2020}%
  \BibitemOpen
  \bibfield  {author} {\bibinfo {author} {\bibfnamefont {A.}~\bibnamefont
  {Gnech}}\ and\ \bibinfo {author} {\bibfnamefont {M.}~\bibnamefont
  {Viviani}},\ }\bibfield  {title} {\bibinfo {title} {Time-reversal violation
  in light nuclei},\ }\href {https://doi.org/10.1103/PhysRevC.101.024004}
  {\bibfield  {journal} {\bibinfo  {journal} {Phys. Rev. C}\ }\textbf {\bibinfo
  {volume} {101}},\ \bibinfo {pages} {024004} (\bibinfo {year}
  {2020})}\BibitemShut {NoStop}%
\bibitem [{\citenamefont {Yang}\ \emph {et~al.}(2020)\citenamefont {Yang},
  \citenamefont {Mereghetti}, \citenamefont {Platter}, \citenamefont
  {Schindler},\ and\ \citenamefont {Vanasse}}]{yang2020electric}%
  \BibitemOpen
  \bibfield  {author} {\bibinfo {author} {\bibfnamefont {Z.}~\bibnamefont
  {Yang}}, \bibinfo {author} {\bibfnamefont {E.}~\bibnamefont {Mereghetti}},
  \bibinfo {author} {\bibfnamefont {L.}~\bibnamefont {Platter}}, \bibinfo
  {author} {\bibfnamefont {M.~R.}\ \bibnamefont {Schindler}},\ and\ \bibinfo
  {author} {\bibfnamefont {J.}~\bibnamefont {Vanasse}},\ }\href@noop {}
  {\bibinfo {title} {Electric dipole moments of three-nucleon systems in the
  pionless effective field theory}} (\bibinfo {year} {2020}),\ \Eprint
  {https://arxiv.org/abs/2011.01885} {arXiv:2011.01885 [nucl-th]} \BibitemShut
  {NoStop}%
\bibitem [{\citenamefont {Yamanaka}\ and\ \citenamefont
  {Hiyama}(2015)}]{Yamanaka2015}%
  \BibitemOpen
  \bibfield  {author} {\bibinfo {author} {\bibfnamefont {N.}~\bibnamefont
  {Yamanaka}}\ and\ \bibinfo {author} {\bibfnamefont {E.}~\bibnamefont
  {Hiyama}},\ }\bibfield  {title} {\bibinfo {title} {Enhancement of the
  $cp$-odd effect in the nuclear electric dipole moment of $^{6}\mathrm{Li}$},\
  }\href {https://doi.org/10.1103/PhysRevC.91.054005} {\bibfield  {journal}
  {\bibinfo  {journal} {Phys. Rev. C}\ }\textbf {\bibinfo {volume} {91}},\
  \bibinfo {pages} {054005} (\bibinfo {year} {2015})}\BibitemShut {NoStop}%
\bibitem [{\citenamefont {Yamanaka}\ \emph {et~al.}(2017)\citenamefont
  {Yamanaka}, \citenamefont {Yamada}, \citenamefont {Hiyama},\ and\
  \citenamefont {Funaki}}]{Yamanaka2017a}%
  \BibitemOpen
  \bibfield  {author} {\bibinfo {author} {\bibfnamefont {N.}~\bibnamefont
  {Yamanaka}}, \bibinfo {author} {\bibfnamefont {T.}~\bibnamefont {Yamada}},
  \bibinfo {author} {\bibfnamefont {E.}~\bibnamefont {Hiyama}},\ and\ \bibinfo
  {author} {\bibfnamefont {Y.}~\bibnamefont {Funaki}},\ }\bibfield  {title}
  {\bibinfo {title} {Electric dipole moment of $^{13}\mathrm{C}$},\ }\href
  {https://doi.org/10.1103/PhysRevC.95.065503} {\bibfield  {journal} {\bibinfo
  {journal} {Phys. Rev. C}\ }\textbf {\bibinfo {volume} {95}},\ \bibinfo
  {pages} {065503} (\bibinfo {year} {2017})}\BibitemShut {NoStop}%
\bibitem [{\citenamefont {Yamanaka}(2017)}]{Yamanaka2017b}%
  \BibitemOpen
  \bibfield  {author} {\bibinfo {author} {\bibfnamefont {N.}~\bibnamefont
  {Yamanaka}},\ }\bibfield  {title} {\bibinfo {title} {Review of the electric
  dipole moment of light nuclei},\ }\href
  {https://doi.org/10.1142/S0218301317300028} {\bibfield  {journal} {\bibinfo
  {journal} {International Journal of Modern Physics E}\ }\textbf {\bibinfo
  {volume} {26}},\ \bibinfo {pages} {1730002} (\bibinfo {year} {2017})},\
  \Eprint {https://arxiv.org/abs/https://doi.org/10.1142/S0218301317300028}
  {https://doi.org/10.1142/S0218301317300028} \BibitemShut {NoStop}%
\bibitem [{\citenamefont {Yamanaka}(2018)}]{Yamanaka2018}%
  \BibitemOpen
  \bibfield  {author} {\bibinfo {author} {\bibfnamefont {N.}~\bibnamefont
  {Yamanaka}},\ }\bibfield  {title} {\bibinfo {title} {Electric dipole moment
  of light nuclei},\ }\href {https://doi.org/10.1007/s10751-018-1510-6}
  {\bibfield  {journal} {\bibinfo  {journal} {Hyperfine Interactions}\ }\textbf
  {\bibinfo {volume} {239}},\ \bibinfo {pages} {35} (\bibinfo {year}
  {2018})}\BibitemShut {NoStop}%
\bibitem [{\citenamefont {Lee}\ \emph {et~al.}(2019)\citenamefont {Lee},
  \citenamefont {Yamanaka},\ and\ \citenamefont {Hiyama}}]{Yamanaka2019a}%
  \BibitemOpen
  \bibfield  {author} {\bibinfo {author} {\bibfnamefont {J.}~\bibnamefont
  {Lee}}, \bibinfo {author} {\bibfnamefont {N.}~\bibnamefont {Yamanaka}},\ and\
  \bibinfo {author} {\bibfnamefont {E.}~\bibnamefont {Hiyama}},\ }\bibfield
  {title} {\bibinfo {title} {Effect of the pauli exclusion principle in the
  electric dipole moment of $^{9}\mathrm{Be}$ with
  $|\mathrm{\ensuremath{\Delta}}s|=1$ interactions},\ }\href
  {https://doi.org/10.1103/PhysRevC.99.055503} {\bibfield  {journal} {\bibinfo
  {journal} {Phys. Rev. C}\ }\textbf {\bibinfo {volume} {99}},\ \bibinfo
  {pages} {055503} (\bibinfo {year} {2019})}\BibitemShut {NoStop}%
\bibitem [{\citenamefont {Yamanaka}\ \emph {et~al.}(2019)\citenamefont
  {Yamanaka}, \citenamefont {Yamada},\ and\ \citenamefont
  {Funaki}}]{Yamanaka2019b}%
  \BibitemOpen
  \bibfield  {author} {\bibinfo {author} {\bibfnamefont {N.}~\bibnamefont
  {Yamanaka}}, \bibinfo {author} {\bibfnamefont {T.}~\bibnamefont {Yamada}},\
  and\ \bibinfo {author} {\bibfnamefont {Y.}~\bibnamefont {Funaki}},\
  }\bibfield  {title} {\bibinfo {title} {Nuclear electric dipole moment in the
  cluster model with a triton: $^{7}\mathrm{Li}$ and $^{11}\mathrm{B}$},\
  }\href {https://doi.org/10.1103/PhysRevC.100.055501} {\bibfield  {journal}
  {\bibinfo  {journal} {Phys. Rev. C}\ }\textbf {\bibinfo {volume} {100}},\
  \bibinfo {pages} {055501} (\bibinfo {year} {2019})}\BibitemShut {NoStop}%
\bibitem [{\citenamefont {Navr\'atil}\ \emph
  {et~al.}(2000{\natexlab{a}})\citenamefont {Navr\'atil}, \citenamefont
  {Vary},\ and\ \citenamefont {Barrett}}]{PhysRevLett.84.5728}%
  \BibitemOpen
  \bibfield  {author} {\bibinfo {author} {\bibfnamefont {P.}~\bibnamefont
  {Navr\'atil}}, \bibinfo {author} {\bibfnamefont {J.~P.}\ \bibnamefont
  {Vary}},\ and\ \bibinfo {author} {\bibfnamefont {B.~R.}\ \bibnamefont
  {Barrett}},\ }\bibfield  {title} {\bibinfo {title} {Properties of ${}^{12}$c
  in the ab initio nuclear shell model},\ }\href
  {https://doi.org/10.1103/PhysRevLett.84.5728} {\bibfield  {journal} {\bibinfo
   {journal} {Phys. Rev. Lett.}\ }\textbf {\bibinfo {volume} {84}},\ \bibinfo
  {pages} {5728} (\bibinfo {year} {2000}{\natexlab{a}})}\BibitemShut {NoStop}%
\bibitem [{\citenamefont {Navr\'atil}\ \emph
  {et~al.}(2000{\natexlab{b}})\citenamefont {Navr\'atil}, \citenamefont
  {Vary},\ and\ \citenamefont {Barrett}}]{PhysRevC.62.054311}%
  \BibitemOpen
  \bibfield  {author} {\bibinfo {author} {\bibfnamefont {P.}~\bibnamefont
  {Navr\'atil}}, \bibinfo {author} {\bibfnamefont {J.~P.}\ \bibnamefont
  {Vary}},\ and\ \bibinfo {author} {\bibfnamefont {B.~R.}\ \bibnamefont
  {Barrett}},\ }\bibfield  {title} {\bibinfo {title} {Large-basis ab initio
  no-core shell model and its application to ${}^{12}$c},\ }\href
  {https://doi.org/10.1103/PhysRevC.62.054311} {\bibfield  {journal} {\bibinfo
  {journal} {Phys. Rev. C}\ }\textbf {\bibinfo {volume} {62}},\ \bibinfo
  {pages} {054311} (\bibinfo {year} {2000}{\natexlab{b}})}\BibitemShut
  {NoStop}%
\bibitem [{\citenamefont {Barrett}\ \emph {et~al.}(2013)\citenamefont
  {Barrett}, \citenamefont {Navr\'atil},\ and\ \citenamefont
  {Vary}}]{Barrett2013}%
  \BibitemOpen
  \bibfield  {author} {\bibinfo {author} {\bibfnamefont {B.~R.}\ \bibnamefont
  {Barrett}}, \bibinfo {author} {\bibfnamefont {P.}~\bibnamefont
  {Navr\'atil}},\ and\ \bibinfo {author} {\bibfnamefont {J.~P.}\ \bibnamefont
  {Vary}},\ }\bibfield  {title} {\bibinfo {title} {Ab initio no core shell
  model},\ }\href {https://doi.org/https://doi.org/10.1016/j.ppnp.2012.10.003}
  {\bibfield  {journal} {\bibinfo  {journal} {Progress in Particle and Nuclear
  Physics}\ }\textbf {\bibinfo {volume} {69}},\ \bibinfo {pages} {131 }
  (\bibinfo {year} {2013})}\BibitemShut {NoStop}%
\bibitem [{\citenamefont {Som\`a}\ \emph {et~al.}(2020)\citenamefont {Som\`a},
  \citenamefont {Navr\'atil}, \citenamefont {Raimondi}, \citenamefont
  {Barbieri},\ and\ \citenamefont {Duguet}}]{PhysRevC.101.014318}%
  \BibitemOpen
  \bibfield  {author} {\bibinfo {author} {\bibfnamefont {V.}~\bibnamefont
  {Som\`a}}, \bibinfo {author} {\bibfnamefont {P.}~\bibnamefont {Navr\'atil}},
  \bibinfo {author} {\bibfnamefont {F.}~\bibnamefont {Raimondi}}, \bibinfo
  {author} {\bibfnamefont {C.}~\bibnamefont {Barbieri}},\ and\ \bibinfo
  {author} {\bibfnamefont {T.}~\bibnamefont {Duguet}},\ }\bibfield  {title}
  {\bibinfo {title} {Novel chiral hamiltonian and observables in light and
  medium-mass nuclei},\ }\href {https://doi.org/10.1103/PhysRevC.101.014318}
  {\bibfield  {journal} {\bibinfo  {journal} {Phys. Rev. C}\ }\textbf {\bibinfo
  {volume} {101}},\ \bibinfo {pages} {014318} (\bibinfo {year}
  {2020})}\BibitemShut {NoStop}%
\bibitem [{\citenamefont {Entem}\ and\ \citenamefont
  {Machleidt}(2003)}]{Entem2003}%
  \BibitemOpen
  \bibfield  {author} {\bibinfo {author} {\bibfnamefont {D.~R.}\ \bibnamefont
  {Entem}}\ and\ \bibinfo {author} {\bibfnamefont {R.}~\bibnamefont
  {Machleidt}},\ }\bibfield  {title} {\bibinfo {title} {{Accurate
  charge-dependent nucleon-nucleon potential at fourth order of chiral
  perturbation theory}},\ }\href {https://doi.org/10.1103/PhysRevC.68.041001}
  {\bibfield  {journal} {\bibinfo  {journal} {Phys. Rev. C}\ }\textbf {\bibinfo
  {volume} {68}},\ \bibinfo {pages} {041001} (\bibinfo {year}
  {2003})}\BibitemShut {NoStop}%
\bibitem [{\citenamefont {Gysbers}\ \emph {et~al.}(2019)\citenamefont
  {Gysbers}, \citenamefont {Hagen}, \citenamefont {Holt}, \citenamefont
  {Jansen}, \citenamefont {Morris}, \citenamefont {Navr{\'a}til}, \citenamefont
  {Papenbrock}, \citenamefont {Quaglioni}, \citenamefont {Schwenk},
  \citenamefont {Stroberg},\ and\ \citenamefont {Wendt}}]{Gysbers2019}%
  \BibitemOpen
  \bibfield  {author} {\bibinfo {author} {\bibfnamefont {P.}~\bibnamefont
  {Gysbers}}, \bibinfo {author} {\bibfnamefont {G.}~\bibnamefont {Hagen}},
  \bibinfo {author} {\bibfnamefont {J.~D.}\ \bibnamefont {Holt}}, \bibinfo
  {author} {\bibfnamefont {G.~R.}\ \bibnamefont {Jansen}}, \bibinfo {author}
  {\bibfnamefont {T.~D.}\ \bibnamefont {Morris}}, \bibinfo {author}
  {\bibfnamefont {P.}~\bibnamefont {Navr{\'a}til}}, \bibinfo {author}
  {\bibfnamefont {T.}~\bibnamefont {Papenbrock}}, \bibinfo {author}
  {\bibfnamefont {S.}~\bibnamefont {Quaglioni}}, \bibinfo {author}
  {\bibfnamefont {A.}~\bibnamefont {Schwenk}}, \bibinfo {author} {\bibfnamefont
  {S.~R.}\ \bibnamefont {Stroberg}},\ and\ \bibinfo {author} {\bibfnamefont
  {K.~A.}\ \bibnamefont {Wendt}},\ }\bibfield  {title} {\bibinfo {title}
  {Discrepancy between experimental and theoretical $\beta$-decay rates
  resolved from first principles},\ }\href
  {https://doi.org/10.1038/s41567-019-0450-7} {\bibfield  {journal} {\bibinfo
  {journal} {Nature Physics}\ }\textbf {\bibinfo {volume} {15}},\ \bibinfo
  {pages} {428} (\bibinfo {year} {2019})}\BibitemShut {NoStop}%
\bibitem [{\citenamefont {Machleidt}\ and\ \citenamefont
  {Entem}(2011)}]{Machleidt2011}%
  \BibitemOpen
  \bibfield  {author} {\bibinfo {author} {\bibfnamefont {R.}~\bibnamefont
  {Machleidt}}\ and\ \bibinfo {author} {\bibfnamefont {D.}~\bibnamefont
  {Entem}},\ }\bibfield  {title} {\bibinfo {title} {Chiral effective field
  theory and nuclear forces},\ }\href
  {https://doi.org/https://doi.org/10.1016/j.physrep.2011.02.001} {\bibfield
  {journal} {\bibinfo  {journal} {Physics Reports}\ }\textbf {\bibinfo {volume}
  {503}},\ \bibinfo {pages} {1 } (\bibinfo {year} {2011})}\BibitemShut
  {NoStop}%
\bibitem [{\citenamefont {H\"uther}\ \emph {et~al.}(2020)\citenamefont
  {H\"uther}, \citenamefont {Vobig}, \citenamefont {Hebeler}, \citenamefont
  {Machleidt},\ and\ \citenamefont {Roth}}]{Huther2020}%
  \BibitemOpen
  \bibfield  {author} {\bibinfo {author} {\bibfnamefont {T.}~\bibnamefont
  {H\"uther}}, \bibinfo {author} {\bibfnamefont {K.}~\bibnamefont {Vobig}},
  \bibinfo {author} {\bibfnamefont {K.}~\bibnamefont {Hebeler}}, \bibinfo
  {author} {\bibfnamefont {R.}~\bibnamefont {Machleidt}},\ and\ \bibinfo
  {author} {\bibfnamefont {R.}~\bibnamefont {Roth}},\ }\bibfield  {title}
  {\bibinfo {title} {Family of chiral two- plus three-nucleon interactions for
  accurate nuclear structure studies},\ }\href
  {https://doi.org/https://doi.org/10.1016/j.physletb.2020.135651} {\bibfield
  {journal} {\bibinfo  {journal} {Physics Letters B}\ }\textbf {\bibinfo
  {volume} {808}},\ \bibinfo {pages} {135651} (\bibinfo {year}
  {2020})}\BibitemShut {NoStop}%
\bibitem [{\citenamefont {Wegner}(1994)}]{Wegner1994}%
  \BibitemOpen
  \bibfield  {author} {\bibinfo {author} {\bibfnamefont {F.}~\bibnamefont
  {Wegner}},\ }\bibfield  {title} {\bibinfo {title} {{Flow-equations for
  Hamiltonians}},\ }\href {https://doi.org/10.1002/andp.19945060203} {\bibfield
   {journal} {\bibinfo  {journal} {Ann. Phys.}\ }\textbf {\bibinfo {volume}
  {506}},\ \bibinfo {pages} {77} (\bibinfo {year} {1994})}\BibitemShut
  {NoStop}%
\bibitem [{\citenamefont {Bogner}\ \emph {et~al.}(2007)\citenamefont {Bogner},
  \citenamefont {Furnstahl},\ and\ \citenamefont {Perry}}]{Bogner2007}%
  \BibitemOpen
  \bibfield  {author} {\bibinfo {author} {\bibfnamefont {S.~K.}\ \bibnamefont
  {Bogner}}, \bibinfo {author} {\bibfnamefont {R.~J.}\ \bibnamefont
  {Furnstahl}},\ and\ \bibinfo {author} {\bibfnamefont {R.~J.}\ \bibnamefont
  {Perry}},\ }\bibfield  {title} {\bibinfo {title} {Similarity renormalization
  group for nucleon-nucleon interactions},\ }\href
  {https://doi.org/10.1103/PhysRevC.75.061001} {\bibfield  {journal} {\bibinfo
  {journal} {Phys. Rev. C}\ }\textbf {\bibinfo {volume} {75}},\ \bibinfo
  {pages} {061001} (\bibinfo {year} {2007})}\BibitemShut {NoStop}%
\bibitem [{\citenamefont {Roth}\ \emph {et~al.}(2008)\citenamefont {Roth},
  \citenamefont {Reinhardt},\ and\ \citenamefont
  {Hergert}}]{PhysRevC.77.064003}%
  \BibitemOpen
  \bibfield  {author} {\bibinfo {author} {\bibfnamefont {R.}~\bibnamefont
  {Roth}}, \bibinfo {author} {\bibfnamefont {S.}~\bibnamefont {Reinhardt}},\
  and\ \bibinfo {author} {\bibfnamefont {H.}~\bibnamefont {Hergert}},\
  }\bibfield  {title} {\bibinfo {title} {Unitary correlation operator method
  and similarity renormalization group: Connections and differences},\ }\href
  {https://doi.org/10.1103/PhysRevC.77.064003} {\bibfield  {journal} {\bibinfo
  {journal} {Phys. Rev. C}\ }\textbf {\bibinfo {volume} {77}},\ \bibinfo
  {pages} {064003} (\bibinfo {year} {2008})}\BibitemShut {NoStop}%
\bibitem [{\citenamefont {Bogner}\ \emph {et~al.}(2010)\citenamefont {Bogner},
  \citenamefont {Furnstahl},\ and\ \citenamefont {Schwenk}}]{Bogner201094}%
  \BibitemOpen
  \bibfield  {author} {\bibinfo {author} {\bibfnamefont {S.}~\bibnamefont
  {Bogner}}, \bibinfo {author} {\bibfnamefont {R.}~\bibnamefont {Furnstahl}},\
  and\ \bibinfo {author} {\bibfnamefont {A.}~\bibnamefont {Schwenk}},\
  }\bibfield  {title} {\bibinfo {title} {From low-momentum interactions to
  nuclear structure},\ }\href
  {https://doi.org/http://dx.doi.org/10.1016/j.ppnp.2010.03.001} {\bibfield
  {journal} {\bibinfo  {journal} {Progress in Particle and Nuclear Physics}\
  }\textbf {\bibinfo {volume} {65}},\ \bibinfo {pages} {94 } (\bibinfo {year}
  {2010})}\BibitemShut {NoStop}%
\bibitem [{\citenamefont {Jurgenson}\ \emph {et~al.}(2009)\citenamefont
  {Jurgenson}, \citenamefont {Navr\'atil},\ and\ \citenamefont
  {Furnstahl}}]{Jurgenson2009}%
  \BibitemOpen
  \bibfield  {author} {\bibinfo {author} {\bibfnamefont {E.~D.}\ \bibnamefont
  {Jurgenson}}, \bibinfo {author} {\bibfnamefont {P.}~\bibnamefont
  {Navr\'atil}},\ and\ \bibinfo {author} {\bibfnamefont {R.~J.}\ \bibnamefont
  {Furnstahl}},\ }\bibfield  {title} {\bibinfo {title} {Evolution of nuclear
  many-body forces with the similarity renormalization group},\ }\href
  {https://doi.org/10.1103/PhysRevLett.103.082501} {\bibfield  {journal}
  {\bibinfo  {journal} {Phys. Rev. Lett.}\ }\textbf {\bibinfo {volume} {103}},\
  \bibinfo {pages} {082501} (\bibinfo {year} {2009})}\BibitemShut {NoStop}%
\bibitem [{\citenamefont {Haxton}\ and\ \citenamefont
  {Henley}(1983)}]{Haxton83}%
  \BibitemOpen
  \bibfield  {author} {\bibinfo {author} {\bibfnamefont {W.~C.}\ \bibnamefont
  {Haxton}}\ and\ \bibinfo {author} {\bibfnamefont {E.~M.}\ \bibnamefont
  {Henley}},\ }\bibfield  {title} {\bibinfo {title} {Enhanced $t$-nonconserving
  nuclear moments},\ }\href {https://doi.org/10.1103/PhysRevLett.51.1937}
  {\bibfield  {journal} {\bibinfo  {journal} {Phys. Rev. Lett.}\ }\textbf
  {\bibinfo {volume} {51}},\ \bibinfo {pages} {1937} (\bibinfo {year}
  {1983})}\BibitemShut {NoStop}%
\bibitem [{\citenamefont {Gudkov}\ \emph {et~al.}(1993)\citenamefont {Gudkov},
  \citenamefont {He},\ and\ \citenamefont {McKellar}}]{Gudkov93}%
  \BibitemOpen
  \bibfield  {author} {\bibinfo {author} {\bibfnamefont {V.~P.}\ \bibnamefont
  {Gudkov}}, \bibinfo {author} {\bibfnamefont {X.-G.}\ \bibnamefont {He}},\
  and\ \bibinfo {author} {\bibfnamefont {B.~H.~J.}\ \bibnamefont {McKellar}},\
  }\bibfield  {title} {\bibinfo {title} {Cp-odd nucleon potential},\ }\href
  {https://doi.org/10.1103/PhysRevC.47.2365} {\bibfield  {journal} {\bibinfo
  {journal} {Phys. Rev. C}\ }\textbf {\bibinfo {volume} {47}},\ \bibinfo
  {pages} {2365} (\bibinfo {year} {1993})}\BibitemShut {NoStop}%
\bibitem [{\citenamefont {Haydock}(1974)}]{Haydock_1974}%
  \BibitemOpen
  \bibfield  {author} {\bibinfo {author} {\bibfnamefont {R.}~\bibnamefont
  {Haydock}},\ }\bibfield  {title} {\bibinfo {title} {The inverse of a linear
  operator},\ }\href {https://doi.org/10.1088/0305-4470/7/17/006} {\bibfield
  {journal} {\bibinfo  {journal} {Journal of Physics A: Mathematical, Nuclear
  and General}\ }\textbf {\bibinfo {volume} {7}},\ \bibinfo {pages} {2120}
  (\bibinfo {year} {1974})}\BibitemShut {NoStop}%
\bibitem [{\citenamefont {Marchisio}\ \emph {et~al.}(2003)\citenamefont
  {Marchisio}, \citenamefont {Barnea}, \citenamefont {Leidemann},\ and\
  \citenamefont {Orlandini}}]{Marchisio2003}%
  \BibitemOpen
  \bibfield  {author} {\bibinfo {author} {\bibfnamefont {M.~A.}\ \bibnamefont
  {Marchisio}}, \bibinfo {author} {\bibfnamefont {N.}~\bibnamefont {Barnea}},
  \bibinfo {author} {\bibfnamefont {W.}~\bibnamefont {Leidemann}},\ and\
  \bibinfo {author} {\bibfnamefont {G.}~\bibnamefont {Orlandini}},\ }\bibfield
  {title} {\bibinfo {title} {Efficient method for lorentz integral transforms
  of reaction cross sections},\ }\href
  {https://doi.org/10.1007/s00601-003-0017-z} {\bibfield  {journal} {\bibinfo
  {journal} {Few-Body Systems}\ }\textbf {\bibinfo {volume} {33}},\ \bibinfo
  {pages} {259} (\bibinfo {year} {2003})}\BibitemShut {NoStop}%
\bibitem [{\citenamefont {Hao}\ \emph {et~al.}(2020)\citenamefont {Hao},
  \citenamefont {Navr\'atil}, \citenamefont {Norrgard}, \citenamefont
  {Ilia\ifmmode~\check{s}\else \v{s}\fi{}}, \citenamefont {Eliav},
  \citenamefont {Timmermans}, \citenamefont {Flambaum},\ and\ \citenamefont
  {Borschevsky}}]{Hao2020}%
  \BibitemOpen
  \bibfield  {author} {\bibinfo {author} {\bibfnamefont {Y.}~\bibnamefont
  {Hao}}, \bibinfo {author} {\bibfnamefont {P.}~\bibnamefont {Navr\'atil}},
  \bibinfo {author} {\bibfnamefont {E.~B.}\ \bibnamefont {Norrgard}}, \bibinfo
  {author} {\bibfnamefont {M.}~\bibnamefont {Ilia\ifmmode~\check{s}\else
  \v{s}\fi{}}}, \bibinfo {author} {\bibfnamefont {E.}~\bibnamefont {Eliav}},
  \bibinfo {author} {\bibfnamefont {R.~G.~E.}\ \bibnamefont {Timmermans}},
  \bibinfo {author} {\bibfnamefont {V.~V.}\ \bibnamefont {Flambaum}},\ and\
  \bibinfo {author} {\bibfnamefont {A.}~\bibnamefont {Borschevsky}},\
  }\bibfield  {title} {\bibinfo {title} {Nuclear spin-dependent
  parity-violating effects in light polyatomic molecules},\ }\href
  {https://doi.org/10.1103/PhysRevA.102.052828} {\bibfield  {journal} {\bibinfo
   {journal} {Phys. Rev. A}\ }\textbf {\bibinfo {volume} {102}},\ \bibinfo
  {pages} {052828} (\bibinfo {year} {2020})}\BibitemShut {NoStop}%
\bibitem [{\citenamefont {Stone}(2005)}]{STONE200575}%
  \BibitemOpen
  \bibfield  {author} {\bibinfo {author} {\bibfnamefont {N.}~\bibnamefont
  {Stone}},\ }\bibfield  {title} {\bibinfo {title} {Table of nuclear magnetic
  dipole and electric quadrupole moments},\ }\href
  {https://doi.org/https://doi.org/10.1016/j.adt.2005.04.001} {\bibfield
  {journal} {\bibinfo  {journal} {Atomic Data and Nuclear Data Tables}\
  }\textbf {\bibinfo {volume} {90}},\ \bibinfo {pages} {75 } (\bibinfo {year}
  {2005})}\BibitemShut {NoStop}%
\bibitem [{\citenamefont {Roth}\ and\ \citenamefont
  {Navr\'atil}(2007)}]{PhysRevLett.99.092501}%
  \BibitemOpen
  \bibfield  {author} {\bibinfo {author} {\bibfnamefont {R.}~\bibnamefont
  {Roth}}\ and\ \bibinfo {author} {\bibfnamefont {P.}~\bibnamefont
  {Navr\'atil}},\ }\bibfield  {title} {\bibinfo {title} {Ab initio study of
  $^{40}\mathrm{Ca}$ with an importance-truncated no-core shell model},\ }\href
  {https://doi.org/10.1103/PhysRevLett.99.092501} {\bibfield  {journal}
  {\bibinfo  {journal} {Phys. Rev. Lett.}\ }\textbf {\bibinfo {volume} {99}},\
  \bibinfo {pages} {092501} (\bibinfo {year} {2007})}\BibitemShut {NoStop}%
\bibitem [{\citenamefont {Roth}(2009)}]{PhysRevC.79.064324}%
  \BibitemOpen
  \bibfield  {author} {\bibinfo {author} {\bibfnamefont {R.}~\bibnamefont
  {Roth}},\ }\bibfield  {title} {\bibinfo {title} {Importance truncation for
  large-scale configuration interaction approaches},\ }\href
  {https://doi.org/10.1103/PhysRevC.79.064324} {\bibfield  {journal} {\bibinfo
  {journal} {Phys. Rev. C}\ }\textbf {\bibinfo {volume} {79}},\ \bibinfo
  {pages} {064324} (\bibinfo {year} {2009})}\BibitemShut {NoStop}%
\bibitem [{\citenamefont {Kwan}\ \emph {et~al.}(2014)\citenamefont {Kwan},
  \citenamefont {Wu}, \citenamefont {Summers}, \citenamefont {Hackman},
  \citenamefont {Drake}, \citenamefont {Andreoiu}, \citenamefont {Ashley},
  \citenamefont {Ball}, \citenamefont {Bender}, \citenamefont {Boston},
  \citenamefont {Boston}, \citenamefont {Chester}, \citenamefont {Close},
  \citenamefont {Cline}, \citenamefont {Cross}, \citenamefont {Dunlop},
  \citenamefont {Finlay}, \citenamefont {Garnsworthy}, \citenamefont {Hayes},
  \citenamefont {Laffoley}, \citenamefont {Nano}, \citenamefont {Navrátil},
  \citenamefont {Pearson}, \citenamefont {Pore}, \citenamefont {Quaglioni},
  \citenamefont {Svensson}, \citenamefont {Starosta}, \citenamefont {Thompson},
  \citenamefont {Voss}, \citenamefont {Williams},\ and\ \citenamefont
  {Wang}}]{KWAN2014210}%
  \BibitemOpen
  \bibfield  {author} {\bibinfo {author} {\bibfnamefont {E.}~\bibnamefont
  {Kwan}}, \bibinfo {author} {\bibfnamefont {C.}~\bibnamefont {Wu}}, \bibinfo
  {author} {\bibfnamefont {N.}~\bibnamefont {Summers}}, \bibinfo {author}
  {\bibfnamefont {G.}~\bibnamefont {Hackman}}, \bibinfo {author} {\bibfnamefont
  {T.}~\bibnamefont {Drake}}, \bibinfo {author} {\bibfnamefont
  {C.}~\bibnamefont {Andreoiu}}, \bibinfo {author} {\bibfnamefont
  {R.}~\bibnamefont {Ashley}}, \bibinfo {author} {\bibfnamefont
  {G.}~\bibnamefont {Ball}}, \bibinfo {author} {\bibfnamefont {P.}~\bibnamefont
  {Bender}}, \bibinfo {author} {\bibfnamefont {A.}~\bibnamefont {Boston}},
  \bibinfo {author} {\bibfnamefont {H.}~\bibnamefont {Boston}}, \bibinfo
  {author} {\bibfnamefont {A.}~\bibnamefont {Chester}}, \bibinfo {author}
  {\bibfnamefont {A.}~\bibnamefont {Close}}, \bibinfo {author} {\bibfnamefont
  {D.}~\bibnamefont {Cline}}, \bibinfo {author} {\bibfnamefont
  {D.}~\bibnamefont {Cross}}, \bibinfo {author} {\bibfnamefont
  {R.}~\bibnamefont {Dunlop}}, \bibinfo {author} {\bibfnamefont
  {A.}~\bibnamefont {Finlay}}, \bibinfo {author} {\bibfnamefont
  {A.}~\bibnamefont {Garnsworthy}}, \bibinfo {author} {\bibfnamefont
  {A.}~\bibnamefont {Hayes}}, \bibinfo {author} {\bibfnamefont
  {A.}~\bibnamefont {Laffoley}}, \bibinfo {author} {\bibfnamefont
  {T.}~\bibnamefont {Nano}}, \bibinfo {author} {\bibfnamefont {P.}~\bibnamefont
  {Navrátil}}, \bibinfo {author} {\bibfnamefont {C.}~\bibnamefont {Pearson}},
  \bibinfo {author} {\bibfnamefont {J.}~\bibnamefont {Pore}}, \bibinfo {author}
  {\bibfnamefont {S.}~\bibnamefont {Quaglioni}}, \bibinfo {author}
  {\bibfnamefont {C.}~\bibnamefont {Svensson}}, \bibinfo {author}
  {\bibfnamefont {K.}~\bibnamefont {Starosta}}, \bibinfo {author}
  {\bibfnamefont {I.}~\bibnamefont {Thompson}}, \bibinfo {author}
  {\bibfnamefont {P.}~\bibnamefont {Voss}}, \bibinfo {author} {\bibfnamefont
  {S.}~\bibnamefont {Williams}},\ and\ \bibinfo {author} {\bibfnamefont
  {Z.}~\bibnamefont {Wang}},\ }\bibfield  {title} {\bibinfo {title} {Precision
  measurement of the electromagnetic dipole strengths in be11},\ }\href
  {https://doi.org/https://doi.org/10.1016/j.physletb.2014.03.049} {\bibfield
  {journal} {\bibinfo  {journal} {Physics Letters B}\ }\textbf {\bibinfo
  {volume} {732}},\ \bibinfo {pages} {210} (\bibinfo {year}
  {2014})}\BibitemShut {NoStop}%
\bibitem [{\citenamefont {Calci}\ \emph {et~al.}(2016)\citenamefont {Calci},
  \citenamefont {Navr\'atil}, \citenamefont {Roth}, \citenamefont
  {Dohet-Eraly}, \citenamefont {Quaglioni},\ and\ \citenamefont
  {Hupin}}]{Calci2016}%
  \BibitemOpen
  \bibfield  {author} {\bibinfo {author} {\bibfnamefont {A.}~\bibnamefont
  {Calci}}, \bibinfo {author} {\bibfnamefont {P.}~\bibnamefont {Navr\'atil}},
  \bibinfo {author} {\bibfnamefont {R.}~\bibnamefont {Roth}}, \bibinfo {author}
  {\bibfnamefont {J.}~\bibnamefont {Dohet-Eraly}}, \bibinfo {author}
  {\bibfnamefont {S.}~\bibnamefont {Quaglioni}},\ and\ \bibinfo {author}
  {\bibfnamefont {G.}~\bibnamefont {Hupin}},\ }\bibfield  {title} {\bibinfo
  {title} {Can ab initio theory explain the phenomenon of parity inversion in
  $^{11}\mathrm{Be}$?},\ }\href
  {https://doi.org/10.1103/PhysRevLett.117.242501} {\bibfield  {journal}
  {\bibinfo  {journal} {Phys. Rev. Lett.}\ }\textbf {\bibinfo {volume} {117}},\
  \bibinfo {pages} {242501} (\bibinfo {year} {2016})}\BibitemShut {NoStop}%
\bibitem [{\citenamefont {de~Vries}\ \emph {et~al.}(2020)\citenamefont
  {de~Vries}, \citenamefont {Epelbaum}, \citenamefont {Girlanda}, \citenamefont
  {Gnech}, \citenamefont {Mereghetti},\ and\ \citenamefont
  {Viviani}}]{deVries2020}%
  \BibitemOpen
  \bibfield  {author} {\bibinfo {author} {\bibfnamefont {J.}~\bibnamefont
  {de~Vries}}, \bibinfo {author} {\bibfnamefont {E.}~\bibnamefont {Epelbaum}},
  \bibinfo {author} {\bibfnamefont {L.}~\bibnamefont {Girlanda}}, \bibinfo
  {author} {\bibfnamefont {A.}~\bibnamefont {Gnech}}, \bibinfo {author}
  {\bibfnamefont {E.}~\bibnamefont {Mereghetti}},\ and\ \bibinfo {author}
  {\bibfnamefont {M.}~\bibnamefont {Viviani}},\ }\bibfield  {title} {\bibinfo
  {title} {Parity- and time-reversal-violating nuclear forces},\ }\href
  {https://doi.org/10.3389/fphy.2020.00218} {\bibfield  {journal} {\bibinfo
  {journal} {Frontiers in Physics}\ }\textbf {\bibinfo {volume} {8}},\ \bibinfo
  {pages} {218} (\bibinfo {year} {2020})}\BibitemShut {NoStop}%
\bibitem [{\citenamefont {de~Vries}\ \emph {et~al.}(2021)\citenamefont
  {de~Vries}, \citenamefont {Gnech},\ and\ \citenamefont {Shain}}]{deVries21}%
  \BibitemOpen
  \bibfield  {author} {\bibinfo {author} {\bibfnamefont {J.}~\bibnamefont
  {de~Vries}}, \bibinfo {author} {\bibfnamefont {A.}~\bibnamefont {Gnech}},\
  and\ \bibinfo {author} {\bibfnamefont {S.}~\bibnamefont {Shain}},\ }\bibfield
   {title} {\bibinfo {title} {Renormalization of $cp$-violating nuclear
  forces},\ }\href {https://doi.org/10.1103/PhysRevC.103.L012501} {\bibfield
  {journal} {\bibinfo  {journal} {Phys. Rev. C}\ }\textbf {\bibinfo {volume}
  {103}},\ \bibinfo {pages} {L012501} (\bibinfo {year} {2021})}\BibitemShut
  {NoStop}%
\bibitem [{\citenamefont {Schuster}\ \emph {et~al.}(2014)\citenamefont
  {Schuster}, \citenamefont {Quaglioni}, \citenamefont {Johnson}, \citenamefont
  {Jurgenson},\ and\ \citenamefont {Navr\'atil}}]{Schuster2014}%
  \BibitemOpen
  \bibfield  {author} {\bibinfo {author} {\bibfnamefont {M.~D.}\ \bibnamefont
  {Schuster}}, \bibinfo {author} {\bibfnamefont {S.}~\bibnamefont {Quaglioni}},
  \bibinfo {author} {\bibfnamefont {C.~W.}\ \bibnamefont {Johnson}}, \bibinfo
  {author} {\bibfnamefont {E.~D.}\ \bibnamefont {Jurgenson}},\ and\ \bibinfo
  {author} {\bibfnamefont {P.}~\bibnamefont {Navr\'atil}},\ }\bibfield  {title}
  {\bibinfo {title} {Operator evolution for ab initio theory of light nuclei},\
  }\href {https://doi.org/10.1103/PhysRevC.90.011301} {\bibfield  {journal}
  {\bibinfo  {journal} {Phys. Rev. C}\ }\textbf {\bibinfo {volume} {90}},\
  \bibinfo {pages} {011301} (\bibinfo {year} {2014})}\BibitemShut {NoStop}%
\bibitem [{\citenamefont {Schuster}\ \emph {et~al.}(2015)\citenamefont
  {Schuster}, \citenamefont {Quaglioni}, \citenamefont {Johnson}, \citenamefont
  {Jurgenson},\ and\ \citenamefont {Navr\'atil}}]{Schuster2015}%
  \BibitemOpen
  \bibfield  {author} {\bibinfo {author} {\bibfnamefont {M.~D.}\ \bibnamefont
  {Schuster}}, \bibinfo {author} {\bibfnamefont {S.}~\bibnamefont {Quaglioni}},
  \bibinfo {author} {\bibfnamefont {C.~W.}\ \bibnamefont {Johnson}}, \bibinfo
  {author} {\bibfnamefont {E.~D.}\ \bibnamefont {Jurgenson}},\ and\ \bibinfo
  {author} {\bibfnamefont {P.}~\bibnamefont {Navr\'atil}},\ }\bibfield  {title}
  {\bibinfo {title} {Operator evolution for ab initio electric dipole
  transitions of $^{4}\mathrm{He}$},\ }\href
  {https://doi.org/10.1103/PhysRevC.92.014320} {\bibfield  {journal} {\bibinfo
  {journal} {Phys. Rev. C}\ }\textbf {\bibinfo {volume} {92}},\ \bibinfo
  {pages} {014320} (\bibinfo {year} {2015})}\BibitemShut {NoStop}%
\bibitem [{\citenamefont {Miyagi}\ \emph {et~al.}(2019)\citenamefont {Miyagi},
  \citenamefont {Abe}, \citenamefont {Kohno}, \citenamefont {Navr\'atil},
  \citenamefont {Okamoto}, \citenamefont {Otsuka}, \citenamefont {Shimizu},\
  and\ \citenamefont {Stroberg}}]{Miyagi2019}%
  \BibitemOpen
  \bibfield  {author} {\bibinfo {author} {\bibfnamefont {T.}~\bibnamefont
  {Miyagi}}, \bibinfo {author} {\bibfnamefont {T.}~\bibnamefont {Abe}},
  \bibinfo {author} {\bibfnamefont {M.}~\bibnamefont {Kohno}}, \bibinfo
  {author} {\bibfnamefont {P.}~\bibnamefont {Navr\'atil}}, \bibinfo {author}
  {\bibfnamefont {R.}~\bibnamefont {Okamoto}}, \bibinfo {author} {\bibfnamefont
  {T.}~\bibnamefont {Otsuka}}, \bibinfo {author} {\bibfnamefont
  {N.}~\bibnamefont {Shimizu}},\ and\ \bibinfo {author} {\bibfnamefont {S.~R.}\
  \bibnamefont {Stroberg}},\ }\bibfield  {title} {\bibinfo {title}
  {Ground-state properties of doubly magic nuclei from the
  unitary-model-operator approach with chiral two- and three-nucleon forces},\
  }\href {https://doi.org/10.1103/PhysRevC.100.034310} {\bibfield  {journal}
  {\bibinfo  {journal} {Phys. Rev. C}\ }\textbf {\bibinfo {volume} {100}},\
  \bibinfo {pages} {034310} (\bibinfo {year} {2019})}\BibitemShut {NoStop}%
\end{thebibliography}
\end{document}